\documentclass[traditabstract]{aa} 

\usepackage{graphicx}
\usepackage{amsmath}
\usepackage{natbib}
\usepackage{epsfig}
\usepackage{txfonts}

\begin{document}

\author{A. Arcones, and H.-T. Janka}

\title{Nucleosynthesis-relevant conditions in neutrino-driven
\\  supernova outflows} 

\subtitle{II. The reverse shock in two-dimensional simulations}

\author{A.~Arcones \inst{1} \and H.-Th. Janka\inst{2}}

\institute{Department of Physics, University of Basel,
  Klingelbergstra{\ss}e 82, CH-4056, Switzerland \and
  Max-Planck-Institut f\"ur Astrophysik, Karl-Schwarzschild-Stra{\ss}e
  1, D-85741 Garching, Germany}

\offprints{A. Arcones} 
\mail{a.arcones@unibas.ch}


\abstract{After the initiation of the explosion of core-collapse
  supernovae, neutrinos emitted from the nascent neutron star drive a
  supersonic baryonic outflow. This neutrino-driven wind interacts
  with the more slowly moving, earlier supernova ejecta forming a wind
  termination shock (or reverse shock), which changes the local wind
  conditions and their evolution. Important nucleosynthesis processes
  (alpha-process, charged-particle reactions, r-process, and
  $\nu$p-process) occur or might occur in this environment.  The
  nucleosynthesis depends on the long-time evolution of density,
  temperature, and expansion velocity. Here we present two-dimensional
  hydrodynamical simulations with an approximate description of
  neutrino-transport effects, which for the first time follow the
  post-bounce accretion, onset of the explosion, wind formation, and
  the wind expansion through the collision with the preceding
  supernova ejecta. Our results demonstrate that the anisotropic
  ejecta distribution has a great impact on the position of the
  reverse shock, the wind profile, and the long-time evolution.  This
  suggests that hydrodynamic instabilities after core bounce and the
  consequential asymmetries may have important effects on the
  nucleosynthesis-relevant conditions in the neutrino-heated baryonic
  mass flow from proto-neutron stars.}


\authorrunning{A.~Arcones and H.-Th.~Janka}  \titlerunning{Two-dimensional
  neutrino-driven winds}

\maketitle

\section{Introduction}
\label{sec:introduction}

Supernova outflows are an important astrophysical site for several
nucleosynthesis processes. A variety of isotopes seem to be
exclusively produced there and contribute to the metal enrichment of
the interstellar medium from which old halo stars and later our Solar
System have formed. Therefore, the fingerprints of supernova
nucleosynthesis are searched for in the old halo stars and are
attempted to be extracted from the Solar System abundances
\cite[e.g.,][]{Sneden.etal:2008}. Studying the long-time evolution of
core-collapse supernovae is a challenging problem because the
explosion mechanism is not yet fully understood
\cite[]{Janka.Langanke.ea:2007} and because of the high computational
demands for simulating such a dynamical environment. CPU time
requirements become even more extreme when multidimensional
simulations combined with accurate neutrino transport are supposed to
follow the supernova ejecta for many seconds. In this paper, we take
advantage of a computationally efficient neutrino transport
approximation \cite[]{Scheck.Kifonidis.Janka.Mueller:2006} to present
the first results of two-dimensional simulations of core-collapse
supernovae that track the outflow evolution for up to three seconds.

Simultaneously with the onset of the core-collapse supernova
explosion, the proto-neutron star forms and cools by emitting
neutrinos \cite[]{Pons.Reddy.ea:1999}. The latter deposit energy in
the layers near the proto-neutron star surface and thus drive a
baryonic mass outflow. Because of ongoing neutrino-energy input, the
entropy of this wind grows with distance from the neutron star, and
the pressure gradient exceeds the gravitational force of the compact
remnant. Therefore the neutrino-heated matter is accelerated quickly
and even reaches supersonic velocities
\cite[]{duncan.shapiro.wasserman:1986}. After first promising results
of \cite{Woosley.Wilson.ea:1994}, neutrino-driven winds have been
investigated intensely as a site where heavy elements could be
produced via rapid neutron capture \cite[see][for a review of
r-process sites and physics]{arnould.goriely.takahashi:2007}. However,
the conditions for successful r-processing found by
\cite{Woosley.Wilson.ea:1994} could neither be confirmed by other
simulations at that time \cite[]{Witti.Janka.Takahashi:1994,
  Takahashi.Witti.Janka:1994}, nor by more recent ones
\cite[e.g.,][]{arcones.janka.scheck:2007,Fischer.etal:2010}. \cite{Roberts.etal:2010}
explain the numerical problems that artificially produced suitable
conditions for the production of heavy n-rich elements
\cite[]{hoffman.woosley.qian:1997} in the simulations of
\cite{Woosley.Wilson.ea:1994}.  Because of the lack of the appropriate
conditions in more recent neutrino-driven wind simulations, efforts
continue to find possible missing physical ingredients. In
\cite{arcones.janka.scheck:2007}, hereafter Paper~I, we studied in
detail the evolution of the neutrino-driven wind and its interaction
with the earlier supernova ejecta for different stellar progenitors
and neutrino luminosities.  This interaction results in a wind
termination shock (or reverse shock), which changes the evolution of
nucleosynthesis-relevant conditions: density, temperature, expansion
velocity (see Paper~I). However, integrated nucleosynthesis based on
those simulations \cite[]{Arcones.Montes:2010} show that no heavy
r-process elements can be produced.

First systematic studies of neutrino-driven winds were based on
solving stationary wind equations \cite[see
e.g.,][]{Qian.Woosley:1996, Otsuki.Tagoshi.ea:2000,
  Thompson.Burrows.Meyer:2001}. These steady-state models could not
consistently account for the reverse shock, which is a hydrodynamical
feature found in several simulations \cite[e.g.,][]{Janka95,
  Janka.Mueller:1996, Burrows.Hayes.Fryxell:1995, Buras.Rampp.ea:2006,
  arcones.janka.scheck:2007, Fischer.etal:2010}. The impact of the
reverse shock on the r-process has been investigated by means of
parametric, steady-state models of subsonic ``breeze'' solutions
combined with an outer boundary at constant pressure
\cite[]{Sumiyoshi00, Terasawa.Sumiyoshi.ea:2002} or supersonic winds
with fixed asymptotic temperature \cite[e.g.,][]{Wanajo.Itoh.ea:2002,
  Wanajo:2007}, or two-stage outflow models where a faster wind phase
is followed by a slower expansion phase
\cite[]{Kuroda.Wanajo.Nomoto:2008, Panov.Janka:2009}. There are also
recent studies \cite[]{Wanajo.etal:2010, Roberts.etal:2010} of the
effect of the reverse shock on the $\nu$p-process
\cite[]{Froehlich.Martinez-Pinedo.ea:2006, Pruet.Hoffman.ea:2006,
  Wanajo:2006}. A discussion of the implications of the reverse shock
on the nucleosynthesis will be given in Sect.~\ref{sec:nuc}.

In Paper~I, we have analyzed the behaviour of the reverse shock based
on spherically symmetric hydrodynamic 
simulations. We found by analytic means that
the position of the reverse shock depends on wind velocity and
density, but also on the pressure of the supernova ejecta with which
the wind collides. This pressure is strongly related to the explosion
energy and progenitor structure. Therefore, the evolution of a mass
element that has crossed the reverse shock is more complicated than a
simple boundary at constant pressure or temperature. Here we show that
multidimensional effects lead to an anisotropic mass and density
distribution of the initial supernova ejecta, which has a big impact
on the position of the reverse shock. The location of the latter
becomes angle dependent. In our two-dimensional (2D) simulations, the
neutrino-driven wind boundary is therefore strongly deformed to a
non-spherical shape. As a consequence, the properties of the shocked
wind medium are strongly dependent on the direction. The results
presented here will help to improve the simple extrapolations for the
ejecta evolution that are used in nucleosynthesis studies. They will
allow to constrain the parameter space of possible wind histories.

The paper is organized as follows. Numerical method and computed
models are described in Sect.~\ref{sec:method}. Our results of two
simulations for a chosen stellar progenitor are presented in
Sect.~\ref{sec:results}, where we also compare 2D to 1D results
(Sect.~\ref{sec:2d1d}) and analyze the impact of varying the
progenitor (Sect.~\ref{sec:prog}). A detailed discussion of
  the assumptions, approximations, and future improvements is included
  in Sect.~\ref{sec:discussion}. Finally, the possible
nucleosynthesis implications of our results are addressed in
Sect.~\ref{sec:nuc} and we summarize our findings in
Sect.~\ref{sec:conclusions}.

\section{Two-dimensional simulations}
\label{sec:method}

Observations indicate that core-collapse supernova are highly
anisotropic. Therefore, multidimensional simulations are more
realistic than spherically symmetric ones. However, they are
computationally much more expensive. Using the same hydro code as in
Paper~I, we have performed hydrodynamical simulations of the
neutrino-driven wind phase in two dimensions. We follow the evolution
of the supernova ejecta for two seconds starting at a few milliseconds
after bounce. As in Paper~I, our simulations are done with a Newtonian
hydrodynamics code, which includes general relativity corrections in
the gravitational potential \cite[]{Marek.Dimmelmeier.ea:2006}. It is
combined with an efficient neutrino transport approximation
\cite[]{Scheck.Kifonidis.Janka.Mueller:2006}. In order to improve the
performance of our simulations, the central high-density region of the
neutron star is excised and its behaviour is prescribed by an inner
boundary condition. This allows us also to vary the contraction and
final radius of the neutrons star, which are determined by the
uncertain high-density equation of state. Neutrino luminosities at the
boundary are chosen such that the explosion energy is around 1.5~B. A
detailed description of the numerical method, initial models, and
boundary treatment can be found in Paper~I,
\cite{Scheck.Kifonidis.Janka.Mueller:2006}, and \cite{Kifonidis03}.

Here we discuss six different models: two (T10-l1-r1 and T10-l2-r1)
are based on a 10.2~$M_\odot$ star (data provided by A.~Heger,
priv. comm.), three (T15-l2-r1, T15-l1-r1, and T15-l1-r0) on a
15~$M_\odot$ progenitor (s15s7b2, \cite{Woosley.Weaver:1995}), and one
(T25-l5-r4) on a 25~$M_\odot$ progenitor (s25a28, \cite{Heger01}). All
progenitors were followed through core collapse by A.~Marek and are
mapped to our 2D code typically 10$\,$ms after bounce. In all models
the subsequent contraction of the neutron star and its cooling
behaviour are described as explained by
\cite{Scheck.Kifonidis.Janka.Mueller:2006} and defined by parameters
$R_{\mathrm f}$ for the final boundary radius and $t_0$ for the
contraction timescale (see Table~\ref{tab:finalresults2d}).  The
neutrino luminosity imposed at the lower grid boundary is
$L_{\nu_e}^{\mathrm{ib}} + L_{\bar{\nu}_e}^{\mathrm{ib}}= $~52.5,
38.6, and 30.3~B/s for the models including ``l1'', ``l2'', and ``l5''
in the name, respectively. For the model names we follow the same
convection as in Paper~I. In Table~\ref{tab:finalresults2d}, we
summarize the values of the input parameters. The initial
configuration of the progenitor model is spherically symmetric.  Since
the code conserves this symmetry, it is necessary to add nonradial
perturbations to some quantity to trigger the growth of nonradial
hydrodynamic instabilities in regions where the conditions allow for
the development of such instabilities.  Following
\cite{Scheck.Kifonidis.Janka.Mueller:2006} we apply seed perturbations
of the velocity field in our simulations, for which we choose
cell-to-cell random variations with a typical amplitude of 0.1~\%.
The resolution of our two-dimensional simulations is around 900 radial
grid points and 180 angular bins. The number of radial zones is
increased depending on the requirements in the outer layers of the
proto-neutron star, where the density gradient is very steep and
neutrinos decouple from matter.

\section{Results}
\label{sec:results}
In this section we describe the explosion and ejecta evolution for
models T10-l1-r1 and T10-l2-r1, give an analytic explanation of the
reverse shock behaviour in Sect.~\ref{sec:analytics}, and compare to
spherically symmetric models (Sect.~\ref{sec:2d1d}) and to the more
massive progenitors (Sect.~\ref{sec:prog}). The model parameters and
overview of the results are presented in
Table~\ref{tab:finalresults2d}: The proto-neutron star contraction is
characterized by its time scale $t_0$ and the final boundary radius
$R_f$. The boundary luminosity ($L_{\nu_e}^{\mathrm{ib}}+
L_{\bar\nu_e}^{\mathrm{ib}}$) is constant during the first second and
later decays as $t^{-3/2}$. The end of the simulation is denoted by
the time $t_{\mathrm{end}}$ given in seconds after
bounce. $M_{\mathrm{bar}}$ is the baryonic mass of the neutron
star. The neutron star radius $R_{\mathrm{ns}}$ is defined as the
location where the density is $10^{11}\,$g$\,$cm$^{-3}$.  $\Delta
E_{\mathrm{tot}}$ is the total energy radiated in neutrinos of all
flavors (measured in bethe [B] = $10^{51}\,$erg),
$L_{\nu_{\mathrm{e}}}$ and $L_{\bar\nu_{\mathrm{e}}}$ are the
luminosities of electron neutrinos and antineutrinos at 500$\,$km,
$\langle\epsilon_{\nu_{\mathrm{e}}}\rangle$ and
$\langle\epsilon_{\bar\nu_{\mathrm{e}}}\rangle$ are the corresponding
mean energies, $E_{\mathrm{exp}}$ is the explosion energy,
$t_{\mathrm{exp}}$ is the post-bounce time when the explosion sets in
(defined as the moment when the energy of expanding postshock matter
exceeds $10^{49}\,$erg).

\begin{table*}[!ht]
  \caption{Parameters and results at one second after bounce (see text).}
  
\vspace{0.5cm}
\setlength\tabcolsep{5pt}
\begin{tabular}{ccccc|cccccccccc}
  \hline
  \hline
  Model & Contraction & $L_{\nu_e}^{\mathrm{ib}}+   L_{\bar\nu_e}^{\mathrm{ib}}$ & Progenitor Mass & $t_{\mathrm{end}}$ & $M_{\mathrm{bar}}$ & $\Delta E_{\mathrm{tot}}$ & $R_{\mathrm{ns}}$ & $L_{\nu_{\mathrm{e}}}$ & $L_{\overline{\nu}_{\mathrm{e}}}$ & $\langle \epsilon_{\nu_{\mathrm{e}}}\rangle$ & $\langle \epsilon_{\overline{\nu}_{\mathrm{e}}}\rangle$ & $E_{\mathrm{exp}}$ & $t_{\mathrm{exp}}$ \\ 
  &  ($R_{\mathrm{f}}$, $\,t_{0}$)  & [B/s]
  & $[M_{\odot}]$   & $[\mathrm{s}]$ & $[M_{\odot}]$ & $[100 \mathrm{B}]$ & $[\mathrm{km}]$ & $[\mathrm{B}/\mathrm{s}]$ & $[\mathrm{B}/\mathrm{s}]$ & $[\mathrm{MeV}]$ & $[\mathrm{MeV}]$ & $[\mathrm{B}]$ & $[\mathrm{s}]$ \\ 
  \hline
  T10-l1-r1 &  9$\,$km;    $\,0.1\,$s & $52.5$ & 10 &  2.8 &   1.261 &   1.305 &   14.82 &   22.97 &   24.63 &   20.51 &   22.10 &   1.457 &   0.153 \\ 
  T10-l2-r1 &  9$\,$km;    $\,0.1\,$s & $38.6$ & 10 &  2.2 &   1.280 &   1.146 &   13.44 &   21.76 &   22.49 &   21.36 &   22.91 &   0.938 &   0.170 \\ 
  T15-l2-r1 &  9$\,$km;    $\,0.1\,$s & $38.6$ & 15 &  1.5 &   1.421 &   1.228 &   12.76 &   22.60 &   23.23 &   22.27 &   23.75 &   1.405 &   0.184 \\ 
  T15-l1-r0  & 8$\,$km;    $\,0.1\,$s & $52.5$ & 15 &  2.0 &   1.393 &   1.460 &   12.79 &   25.53 &   26.45 &   22.68 &   24.04 &   1.364 &   0.156 \\ 
  T15-l1-r1  & 9$\,$km;    $\,0.1\,$s & $52.5$ & 15 &  1.0 &   1.388 &   1.461 &   13.27 &   27.66 &   28.22 &   22.55 &   23.87 &   1.341 &   0.162 \\ 
  T25-l5-r4  & 10.5$\,$km; $\,0.1\,$s & $30.3$ & 25 &  1.6 &   1.869 &   2.233 &   13.41 &   36.46 &   39.92 &   24.31 &   25.51 &   3.674 &   0.197 \\ 
  \hline
\end{tabular}
\label{tab:finalresults2d}
\end{table*}

\begin{figure*}[!ht]
  \centering
   \begin{tabular}{lr}
    \includegraphics[width=0.45\linewidth]{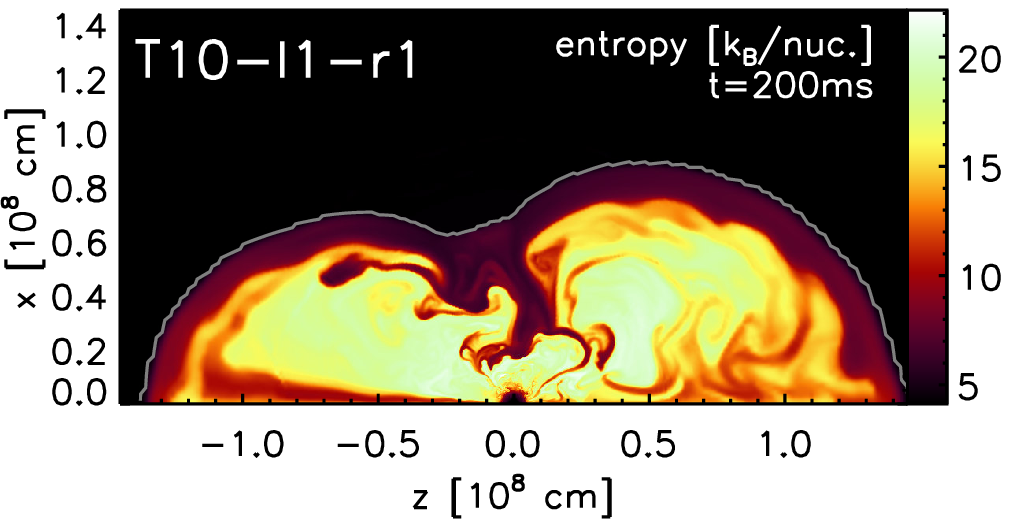} &
    \includegraphics[width=0.45\linewidth]{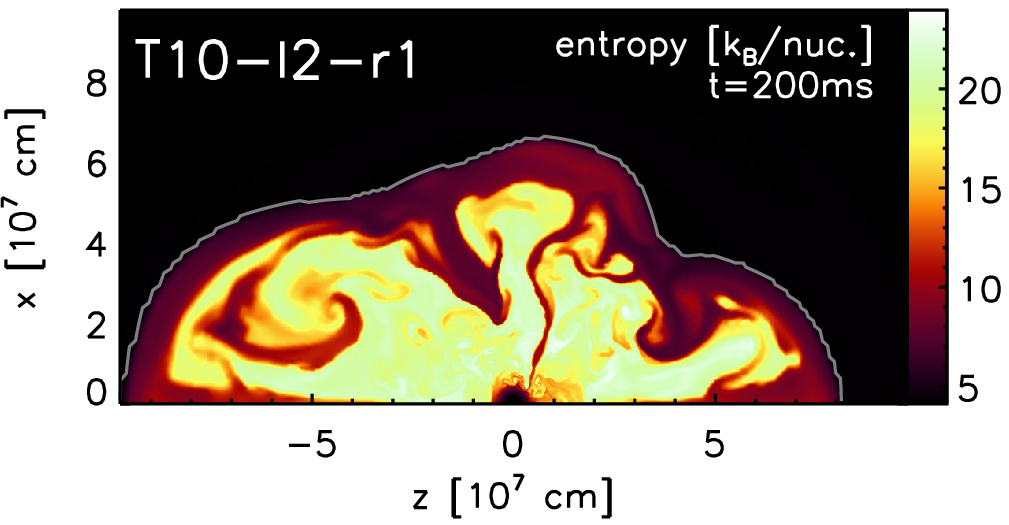} \\
    \includegraphics[width=0.45\linewidth]{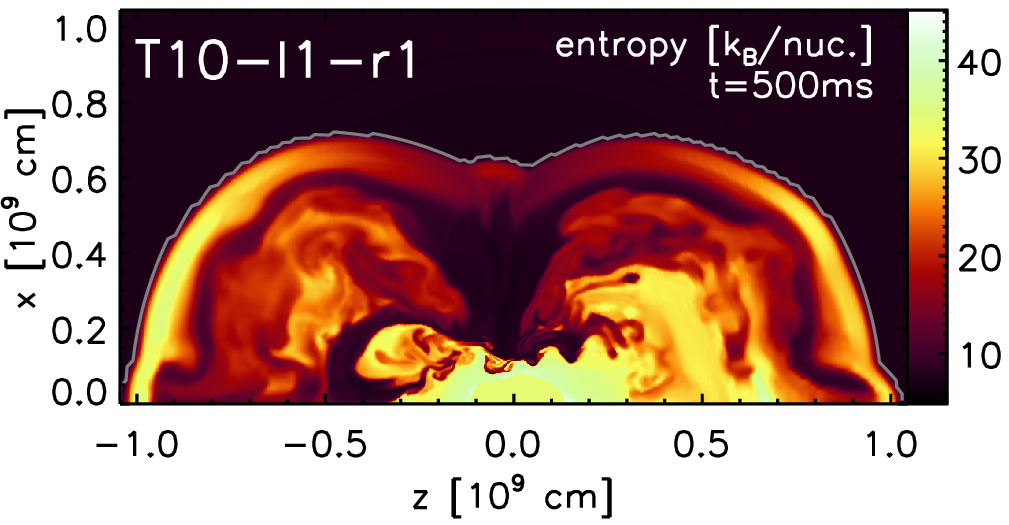} &
    \includegraphics[width=0.45\linewidth]{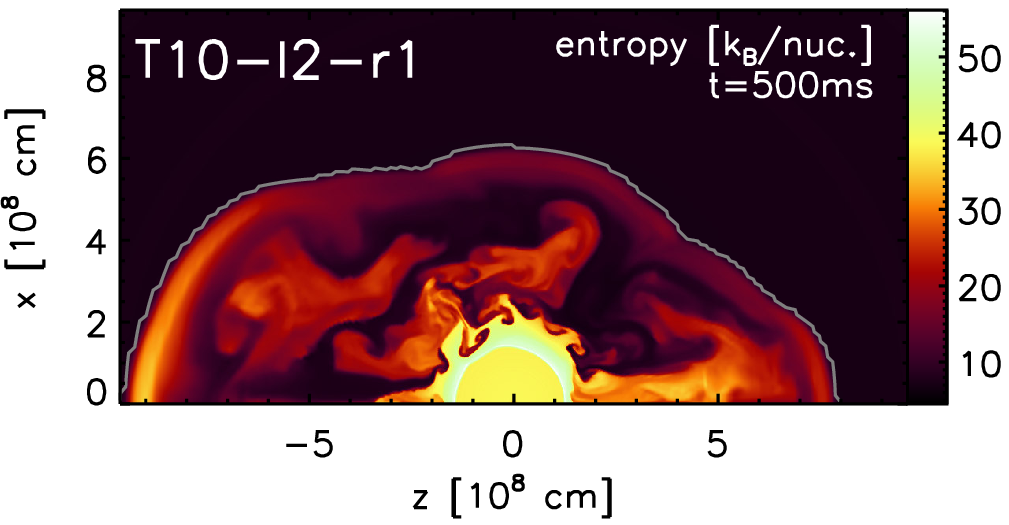}\\
    \includegraphics[width=0.45\linewidth]{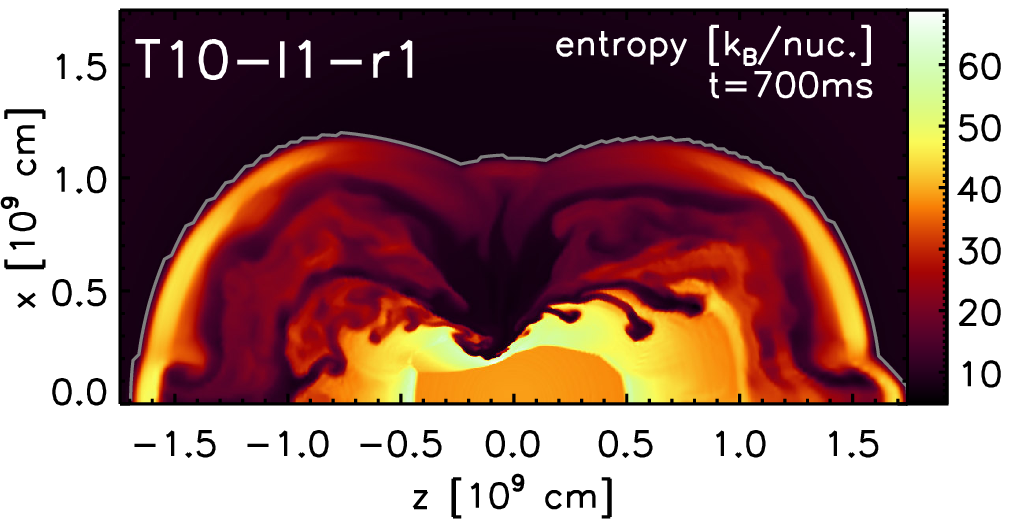} &
    \includegraphics[width=0.45\linewidth]{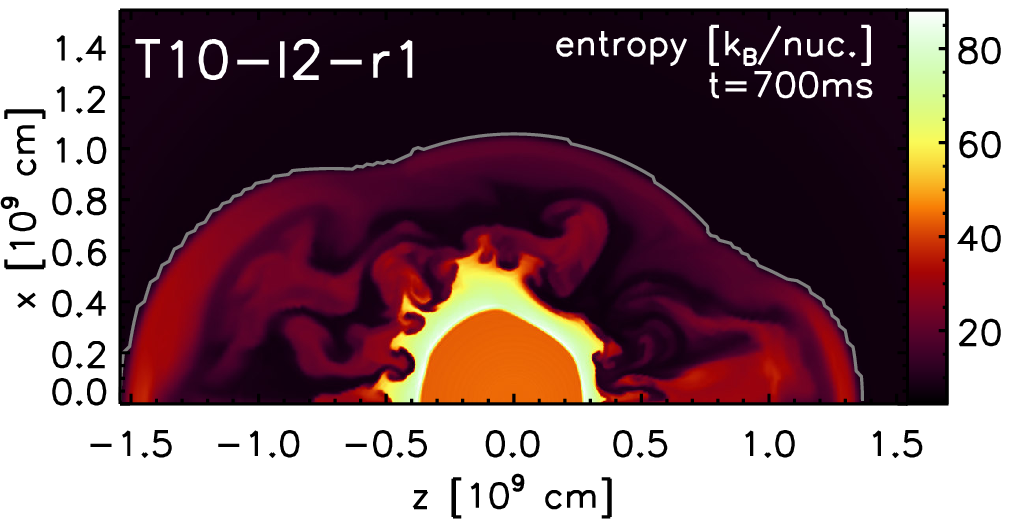}\\ 
    \includegraphics[width=0.45\linewidth]{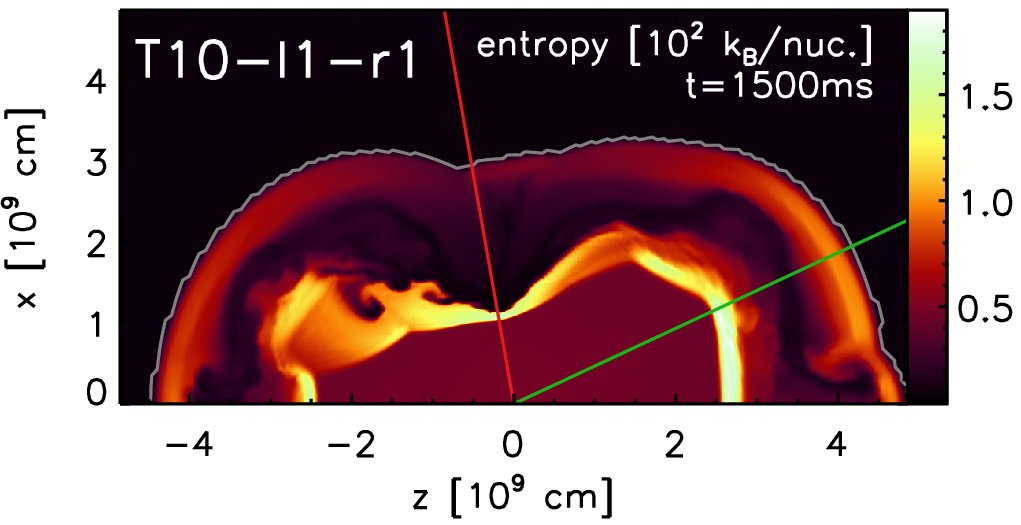}&
    \includegraphics[width=0.45\linewidth]{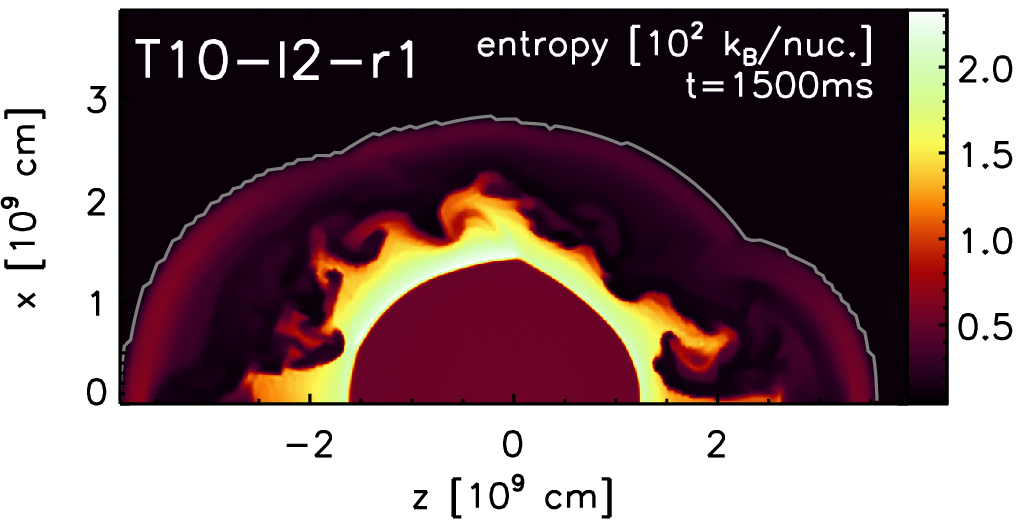}\\
    \includegraphics[width=0.45\linewidth]{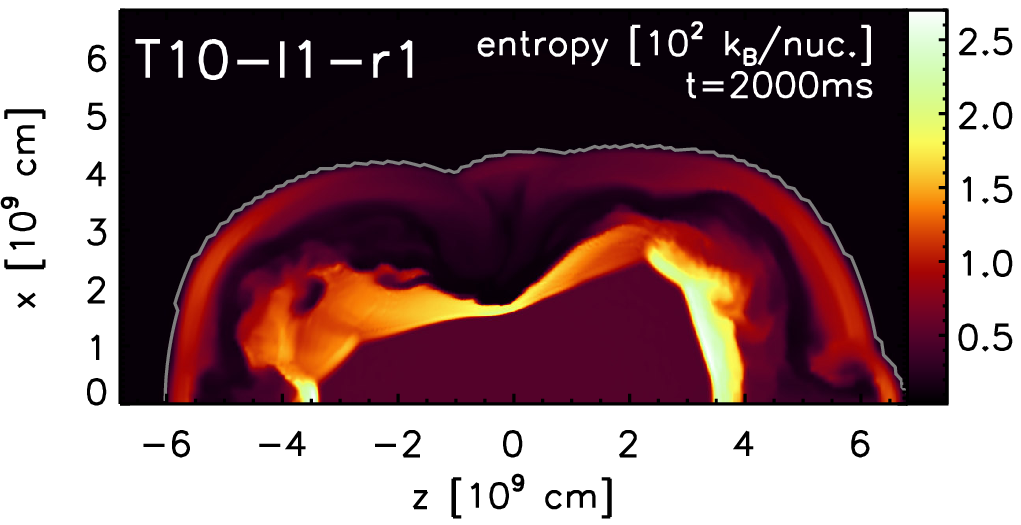}&
    \includegraphics[width=0.45\linewidth]{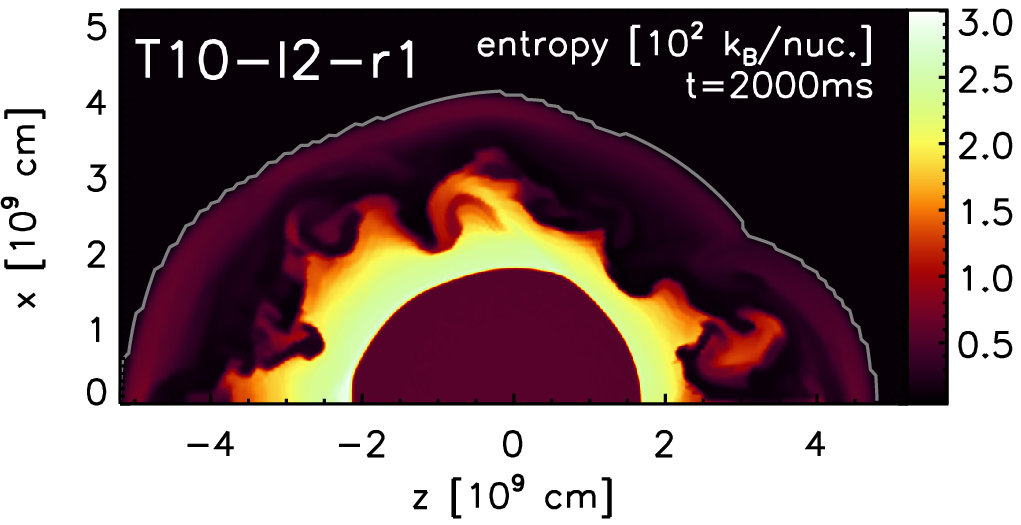}
   \end{tabular}
   \caption{Entropy distribution in models T10-l1-r1 (left column) and
     T10-l2-r1 (right column) for different times after bounce as
     indicated in every panel. The figures are plotted such that the
     polar axis is oriented horizontally with ``south'' ($\theta=\pi$)
     on the left and ``north'' ($\theta=0$) on the right. The thin
     grey line marks the shock radius. In the panel for $t=1500$~ms of
     model T10-l1-r1, the radial lines mark the angular directions
     at $\theta=$~25 degrees 
     (green line) and 100 degrees (red line), along which radial
     profiles are shown in Fig.~\ref{fig:cmp2d-2rad}.}
  \label{fig:T10-stot}
\end{figure*}

The initial distribution is spherically symmetric except for small
random seed perturbations. As neutrinos deposit energy behind the
shock, a negative entropy gradient establishes. The region between
neutron star and shock thus becomes Ledoux-unstable and convective
overturn appears. Neutrino-heated, high entropy matter streams
upwards, while downflows transport low entropy matter from the shock
to the neutron star. The down-flows and rising bubbles evolve quickly
on time scales of the order of tens of milliseconds. The mass
distribution below the shock becomes highly anisotropic and is
dominated by low spherical harmonics modes. In this phase continuous
neutrino heating aided by convection leads to an explosion. The
evolution after the onset of the explosion is shown in
Fig.~\ref{fig:T10-stot} by the entropy distribution at different times
after bounce.

After the launch of the explosion a proto-neutron star forms at the
center and cools and deleptonizes by emitting neutrinos. During this
phase, neutrino-heated matter expands away from the proto-neutron star
surface, in what is known as neutrino-driven wind and collides with the
previous, more slowly moving ejecta. The interaction of the supersonic
wind with the supernova ejecta results in a wind termination shock
(reverse shock). At this discontinuity, kinetic energy is transformed
into internal energy, which produces an increase of the entropy (see
panels of Fig.~\ref{fig:T10-stot} at $t \geq 500$~ms) and the wind is
decelerated. At late times the changes become slower, shock and ejecta
expand quasi-self-similarly.

The neutrino-driven wind is spherically symmetric because it depends
only on the neutrino emission of the proto-neutron star, which is
isotropic. However, the distribution of the early ejecta is highly
anisotropic. This produces a deformation of the reverse shock and of
the shocked wind. 
The reverse shock radius and the properties of the shocked
matter become angle dependent. An example of a two-dimensional feature
is the downflow present at $\theta \approx \pi/2$ in model T10-l1-r1,
which corresponds to the low entropy region visible in the panels for
$t=500-2000$~ms of Fig.~\ref{fig:T10-stot}. This feature leads to a
big deformation of the reverse shock, in contrast to model T10-l2-r1,
where the reverse shock stays almost spherically symmetric as no
strong, long-lasting downflows have formed. These variations in the
evolution and structure of the two models are not an immediate
consequence of the different boundary luminosities. In
Table~\ref{tab:finalresults2d} one can see that many parameters of
both models are very similar. The anisotropy of the ejecta depends on
chaotic effects that are triggered by the initial random perturbations
\cite[for more details see][]{Scheck.Kifonidis.Janka.Mueller:2006}.

High-density, low-entropy regions in the ejecta, which are the
remainders of former downflows, act like obstacles preventing faster
wind expansion in those directions.  In the analytic discussion of
Paper~I, we found that the reverse shock radius ($R_{\mathrm{rs}}$)
depends on the pressure of the more slowly moving early ejecta:
\begin{equation}
  R_{\mathrm{rs}} \propto \sqrt{\frac{\dot{M}_{\mathrm{w}} u_{\mathrm{w}}}{P_{\mathrm{rs}}}} \, .
  \label{eq:Rrs}
\end{equation}
The mass outflow rate ($\dot{M}_{\mathrm{w}}$) and the velocity
($u_{\mathrm{w}}$) of the wind are the same at all angles because the
wind is spherically symmetric. Therefore, the variation of the reverse
shock radius with angle is caused by the pressure variations of the
anisotropic ejecta ($P_{\mathrm{rs}}$).

\begin{figure}[!h]
  \centering
  \includegraphics[width=0.95\linewidth]{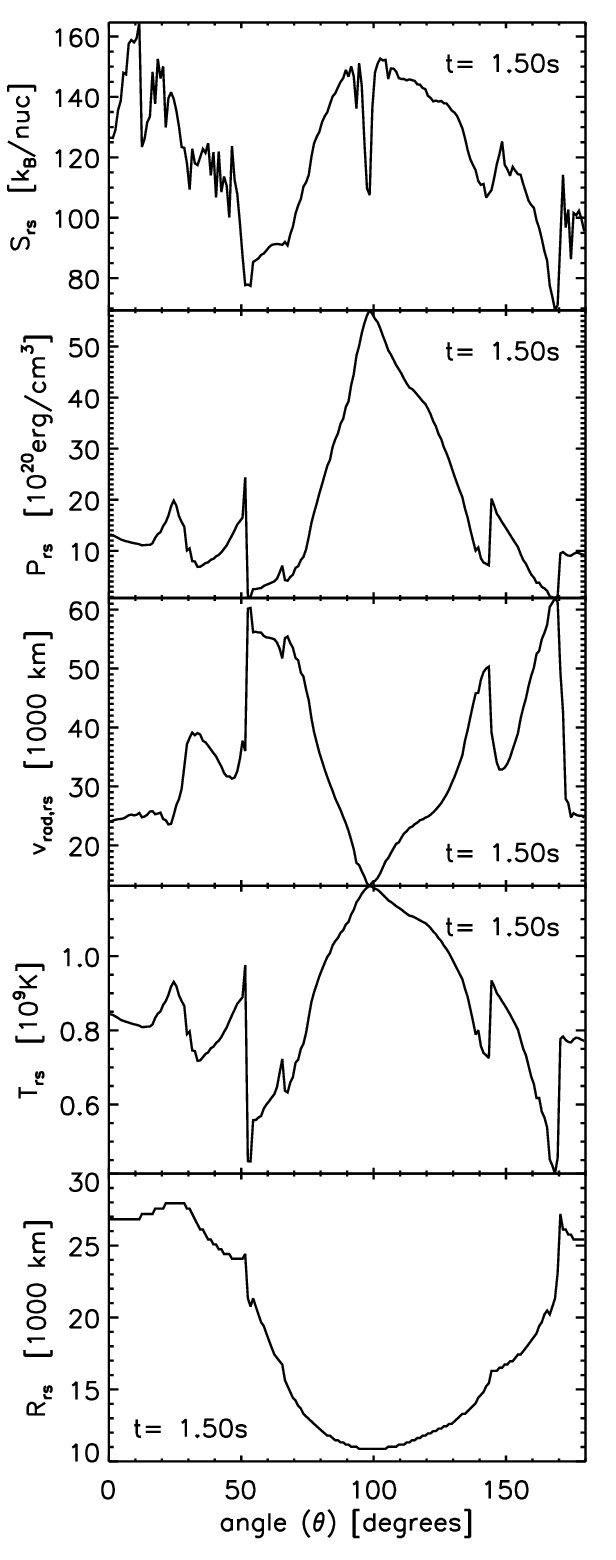}
  \caption{Latitudinal variation of the reverse shock radius (bottom)
    and of the entropy, pressure, radial velocity, and temperature of
    the wind matter just after passing the reverse shock in model
    T10-l1-r1 at $t=1.5$~s after bounce.}
  \label{fig:T10-rs-ang}
\end{figure}

The aspherical matter distribution in the layer between reverse shock
and forward shock leads also to an angle dependence of the entropy
jump, which depends on the reverse-shock position and wind velocity
(approximately as $s_{\mathrm{rs}} \propto \sqrt{R_{\mathrm{rs}}}
u_{\mathrm{w}}^{1.75}$,
\cite{arcones.janka.scheck:2007}). Figure~\ref{fig:T10-rs-ang} shows
the radius of the reverse shock (bottom panel) and the entropy,
pressure, radial velocity, and temperature of the decelerated wind
matter just after the reverse shock as functions of the azimuthal
angle for model T10-l1-r1 at $t=1.5$~s after bounce. Pressure and
reverse shock radius are related as roughly given by
Eq.~(\ref{eq:Rrs}). Moreover, the pressure determines also the angle
between radial direction and reverse shock
(Sect.~\ref{sec:analytics}). The temperature, which is one of the most
relevant quantities for nucleosynthesis, depends mainly on the
position of the reverse shock. When the reverse shock is at a smaller
radius the shocked matter reaches higher temperatures. The link
between these quantities and the geometrical structure of the reverse
shock can be explained by analytic means as we show in the next
section.

\subsection{Analytic discussion}
\label{sec:analytics}

\begin{figure}[!ht]
  \centering
    \includegraphics[width=0.7\linewidth]{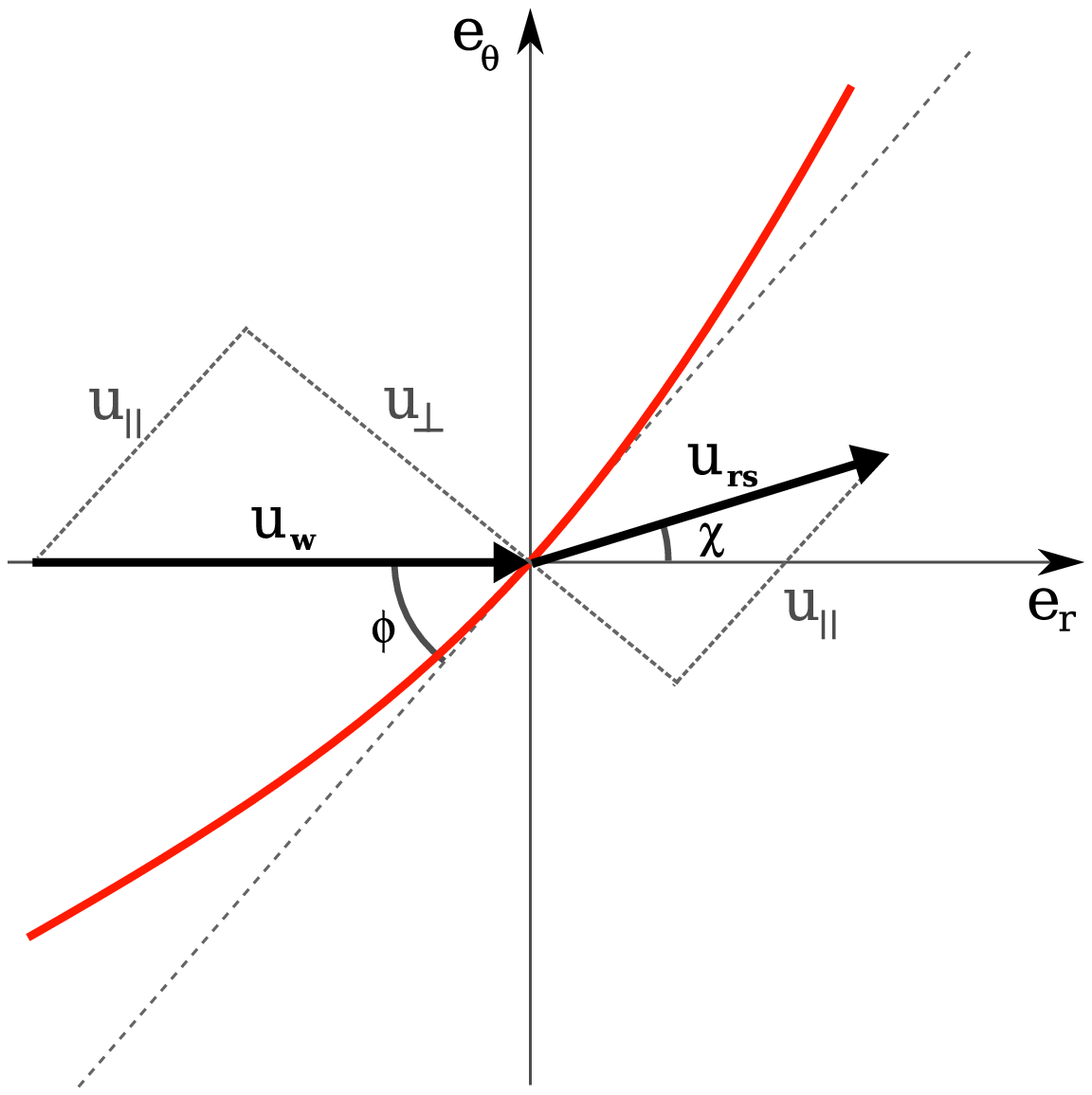} 
    \caption{Schematic representation of the velocities in a fluid
      going through an oblique shock (red line). Only the velocity
      component perpendicular to the shock, $u_{\perp}$, is changed
      when a mass element crosses the shock; the tangential
      component, $u_{\parallel}$, is conserved. Therefore, the
      direction of the flow is changed at the shock.}
  \label{fig:analytic2d}
\end{figure}
We use the Rankine-Hugoniot conditions (see paper~I for further
discussion) to understand the behaviour of the shocked matter and its
angular dependence.  The neutrino-driven wind expands in the radial
direction with velocity $u_{\mathrm{w}}$, hits the slow-moving,
anisotropic ejecta, and a deformed reverse shock results. As shown
schematically in Fig.~\ref{fig:analytic2d}, there is an angle $\phi$
between reverse shock and radial direction. For spherical explosions
this angle is always $\pi/2$, since the reverse shock can only be
perpendicular to the radius vector. In an oblique shock the velocity
can be decomposed into two components: tangential ($u_{\parallel}$)
and perpendicular ($u_{\perp}$) to the shock. The tangential component
of the velocity is continuous through an oblique shock
\cite[]{Landau}. Following the notation introduced in
Fig.~\ref{fig:analytic2d}, this implies that:
\begin{equation}
  u_{\parallel} =  u_{\mathrm{w}} \cos \phi \, =\, u_{\mathrm{rs,r}} \cos \phi +
  u_{\mathrm{rs,\theta}} \sin \phi 
\end{equation}
with
\begin{align}
  u_{\mathrm{rs, r}}     &= u_{\mathrm{rs}} \cos \chi    \, , \\
  u_{\mathrm{rs,\theta}} &= u_{\mathrm{rs}} \sin \chi    \, ,  
\end{align}
being the radial and lateral components of the velocity of the shocked
matter ($u_{\mathrm{rs}}$), respectively.

\begin{figure}[!ht]
  \centering
    \includegraphics[width=0.95\linewidth]{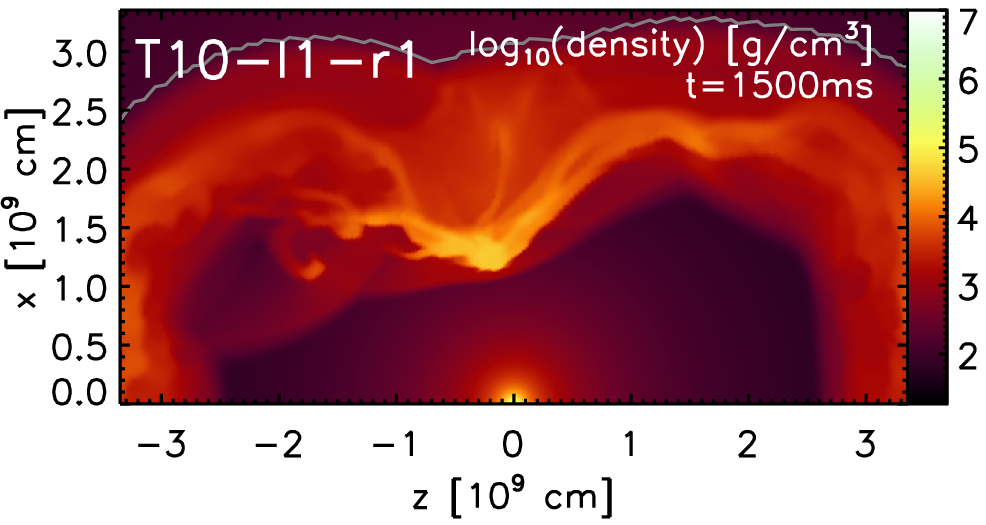} \\
    \includegraphics[width=0.95\linewidth]{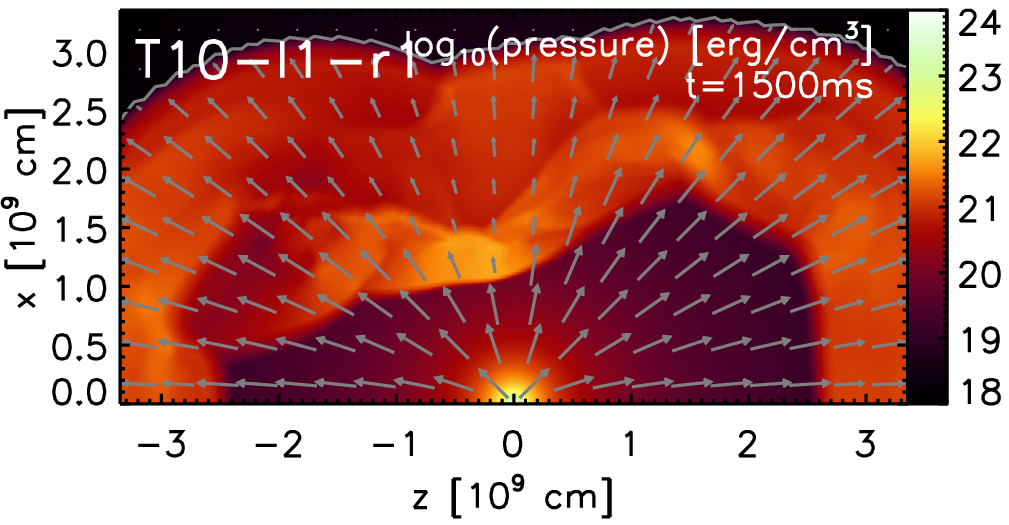} \\
    \includegraphics[width=0.95\linewidth]{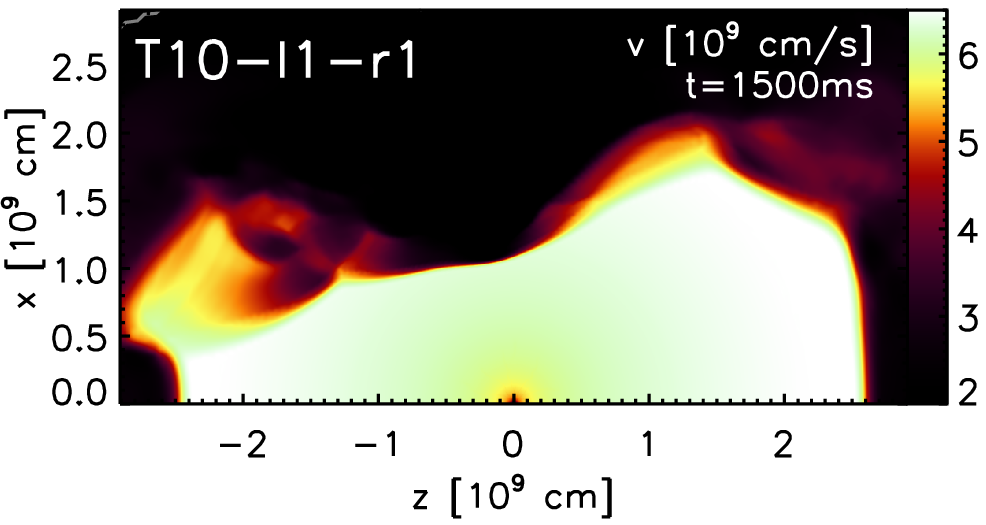} \\
   \caption{Distribution of density, pressure, and absolute value of
     the velocity in the wind and shocked matter for model T10-l1-r1
     at $t=1.5$~s after bounce. In the middle panel the radial
     velocity field is indicated by arrows.}
  \label{fig:T10-denspressvel}
\end{figure}

The perpendicular component of the velocity, $u_{\perp}$, is changed
according to the Rankine-Hugoniot conditions. The mass conservation
condition, including the angle dependence, is:

\begin{equation}
  \rho_{\mathrm{w}} u_{\mathrm{w}} \sin \phi \ = \
  \rho_{\mathrm{rs}} u_{\mathrm{rs,r}} \sin \phi \ + \
  \rho_{\mathrm{rs}} u_{\mathrm{rs,\theta}} \cos \phi \, .  \label{eq:shjump1-2d} 
\end{equation}
The momentum continuity condition for an oblique shock is:
\begin{equation}
  P_{\mathrm{w}}  + \rho_{\mathrm{w}}  u_{\perp}^{2} =
  P_{\mathrm{rs}} + \rho_{\mathrm{rs}} u_{\mathrm{rs,\perp}}^{2} \, ,
  \label{eq:shjump2-2d}
\end{equation}
where only the perpendicular components of the velocities enter. We
can thus write a relation between the pressure and the angle $\phi$:
\begin{equation}
  \Delta P \ = \ \rho_{\mathrm{w}} u_{\mathrm{w}}^{2} \sin^{2} \phi
  \left(1 - \frac{1}{\beta} \right) \, ,
  \label{eq:shjump2-2db}
\end{equation}
where $\Delta P=P_{\mathrm{rs}}-P_{\mathrm{w}}$ is the pressure jump
at the reverse shock (note that usually $P_{\mathrm{w}} \ll
P_{\mathrm{rs}}$), $u_{\mathrm{w}}$ is the radial wind velocity
(Fig.~\ref{fig:analytic2d}), and $\beta$ is:
\begin{equation}
  \beta = 
          \frac{\rho_{\mathrm{rs}}}{\rho_{\mathrm{w}}} 
      \	= \
          \frac{u_{\perp}}{u_{\mathrm{rs},\perp}} 
      \ = \ 
          \frac{ u_{\mathrm{w}} \sin \phi  } 
	       { u_{\mathrm{rs,r}} \sin \phi +u_{\mathrm{rs,\theta}}
                 \cos \phi }\, .
\label{eq:beta2d}
\end{equation}
The variation of the pressure jump (and thus of the entropy and
temperature increase by the reverse shock) with angle $\phi$ can be
deduced from Eq.~(\ref{eq:shjump2-2db}). Since $\beta >1$ usually,
this equation implies that when $\phi$ goes to $\pi/2$, i.e., the
reverse shock is perpendicular to wind velocity, $\Delta P$ and the
entropy jump are larger (see middle panel in
Fig.~\ref{fig:T10-denspressvel}).  As there is (roughly) a pressure
balance between $P_{\mathrm{rs}}$ and the pressure of the slow-moving
early ejecta, which is higher in the downflows, $\phi$ tends to be
about $\pi/2$ in regions where downflows are present. In
Fig.~\ref{fig:T10-denspressvel} the upper panel shows that the
density reaches highest values where the downflow is located
($\theta \approx 100$~degrees). Consistently, the pressure in this
region is also large as shown in the middle panel. This leads to 
straight sections in the reverse shock shape and to the occurrence
of kinks between locations of downflows.

\begin{figure}[!ht]
  \centering
  \includegraphics[width=0.6\linewidth]{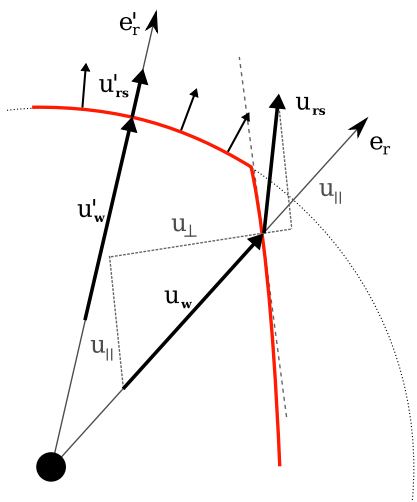}
  \caption{Schematic representation of the velocities in a fluid going
    through the reverse shock (red line) when a kink leads to the
    collimation of the outflowing matter.}
  \label{fig:kink}
\end{figure}

The reverse shock in model T10-l1-r1 thus exhibits several kinks
(Fig.~\ref{fig:T10-denspressvel}) due to the anisotropic pressure
distribution of the ejecta. Figure~\ref{fig:kink} shows, in a
simplified way, the effects of these kinks. An oblique reverse shock
is less effective in decelerating the flow (Eq.~(\ref{eq:beta2d}))
because the tangential component of the velocity is conserved through
the shock.  This leads to higher velocities outside the non-spherical
parts of the reverse shock ($u_{\mathrm{rs}}$ in Fig.~\ref{fig:kink})
compared to spherical regions ($u'_{\mathrm{rs}}$ in
Fig.~\ref{fig:kink}). This can be seen in the velocity field marked by
arrows in the middle panel of Fig.~\ref{fig:T10-denspressvel}. The
velocity of the ejecta is not directed radially after the wind has
passed an oblique shock. This produces collimated high-velocity
outflows starting at the kinks (Fig.~\ref{fig:T10-denspressvel},
bottom panel and Fig.~\ref{fig:T10-stot}). 
Moreover, the kinks of the reverse shock are
associated with minimum values of the pressure and entropy for the
shocked material as visible in Fig.~\ref{fig:T10-rs-ang}.

\subsection{Two-dimensional versus one-dimensional simulations}
\label{sec:2d1d}

Convection enhances neutrino heating, leading to earlier and more
energetic explosions in 2D than in 1D, for the same inner-boundary
parameters. Earlier explosions imply that the neutron star accretes
matter during less time, thus its mass is slightly smaller in 2D (see
Table~\ref{tab:finalresults2d}). These differences alter the wind and
reverse shock properties, making it difficult to contrast exactly one-
and two-dimensional simulations. In this section we compare the
one-dimensional model M10-l1-r1 (see Paper~I) to the two-dimensional
model with the same inner boundary parameters: T10-l1-r1.  First, we
examine the differences between the radial profiles of both models at
a given time and later the evolution of relevant quantities.

\begin{figure}[!ht]
  \centering
  \includegraphics[width=0.97\linewidth]{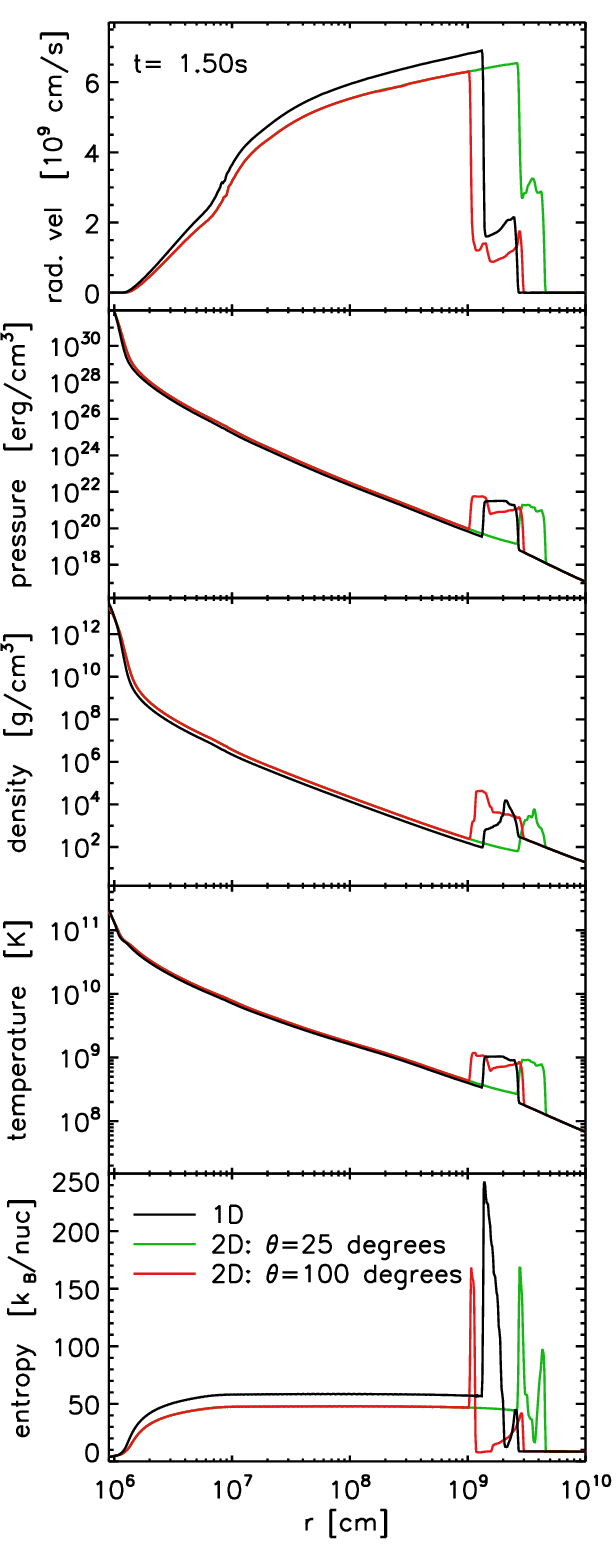}
  \caption{Profiles of radial velocity, pressure, density,
    temperature, and entropy as functions of radius at time 1.5~s
    after bounce. The one-dimensional model M10-l1-r1 (black lines) is
    compared to profiles of the two-dimensional model T10-l1-r1 at
    angles $\theta=25$~degrees (green lines) and $\theta=100$~degrees
    (red lines).}
  \label{fig:cmp2d-2rad}
\end{figure}

Figure~\ref{fig:cmp2d-2rad} shows the radial profiles of different
quantities at 1.5~s after bounce. The one-dimensional model M10-l1-r1
is displayed by black lines. For the two-dimensional model, T10-l1-r1,
the radial profiles differ at all azimuthal angles. Therefore, we
chose two angles (see radial lines in panel for $t=1500$~ms of
Fig.~\ref{fig:T10-stot}): $\theta=25$~degrees (green line) and
$\theta=100$~degrees (red line), where $R_{\mathrm{rs}}$ adopts
extreme values.  As in Paper~I, the radial profiles show an increase
of the entropy in the region where neutrinos deposit energy ($r
\lesssim 20$~km) and a constant entropy value in the wind. Matter
close to the neutron star absorbs neutrinos and moves outwards. The
fast drop of the density and pressure allows the expansion to become
supersonic. The wind velocity increases first approximately as $u
\propto r$ and then approaches an asymptotic value
($u_{\mathrm{w}}$). The interaction of the fast wind with the
slow-moving earlier ejecta results in a sudden drop of the velocity
and a jump of the pressure, density, temperature, and entropy to
higher values at the reverse shock. While the wind is still very
similar in the one- and two-dimensional simulations, there are
significant differences in the conditions of the slow-moving ejecta
and therefore in the reverse shock position. The variations in the
wind are just a consequence of different neutron star properties due
to the evolution towards explosion, which is not the same in one and
two dimensions, as explained before. However, the two profiles of
model T10-l1-r1 are identical in the wind phase indicating that this
region is spherically symmetric. This is expected because the neutrino
emission is isotropic and the neutron star stays spherical without
rotation. The differences in the profiles appear at the position of
the reverse shock, which in 2D depends on the angle and is also
different from the 1D case. Notice that for $\theta=100$~degrees the
velocity of the slow-moving ejecta is very small because of the
presence of relics of a strong downflow.

\begin{figure}[!ht]
  \centering
    \includegraphics[width=0.97\linewidth]{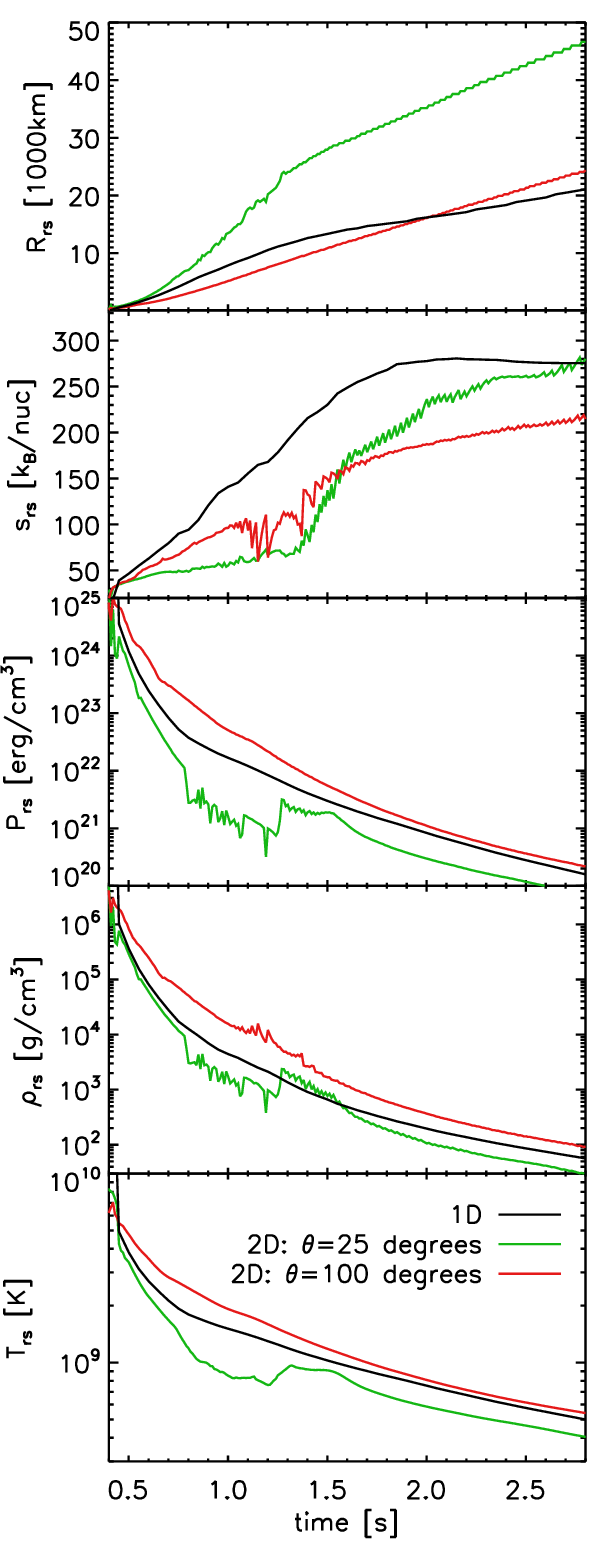} 
    \caption{Evolution of reverse shock radius, and of the entropy,
      pressure, density, and temperature of the shocked neutrino-wind
      matter for models M10-l1-r1 (black lines) and T10-l1-r1 at
      $\theta=25$~degrees (green lines) and $\theta=100$~degrees (red
      lines).}
  \label{fig:rshk_T10}
\end{figure}

Since the main differences between 1D and 2D originate from the reverse
shock radius and are linked to the properties of the supernova ejecta,
we show in Fig.~\ref{fig:rshk_T10} the evolution of the quantities of
matter that has passed the reverse shock. The evolution in 2D is
presented for fixed azimuthal angles $\theta=$~25~degrees (green
lines) and 100~degrees (red lines), corresponding to the lines of same
colors in Fig.~\ref{fig:T10-stot}. In 2D the reverse shock radius
changes significantly between different angles with corresponding
variations of the temperature. This has an impact on nucleosynthesis
as we will discuss later in Sect.~\ref{sec:nuc}. The behaviour of the
reverse shock radius follows Eq.~(\ref{eq:Rrs}).  The entropy in the
two-dimensional simulation is always lower, even when the reverse
shock radius is larger than in the one-dimensional case.  This is
explained by the fact that the reverse shock for the green ray 
is oblique to the radial direction and the entropy jump is thus reduced 
(see Sect.~\ref{sec:analytics}). Moreover, in 2D the neutron-star mass
is slightly smaller, which leads to lower entropies in the wind itself
\cite[]{Qian.Woosley:1996}.

Note that the wind phase can be studied just by one-dimensional
simulations because (without rotation) it stays spherically
symmetric. Even simple steady-state wind models
\cite{Otsuki.Tagoshi.ea:2000, Thompson.Burrows.Meyer:2001} are
sufficient to determine the wind properties, e.g. for discussing the
nucleosynthesis in the wind. However, the interaction of the wind with
the supernova ejecta is a hydrodynamical problem that requires
supernova explosion simulations. Moreover, we have shown here that
one-dimensional models are not enough to account for all the
possibilities of the long-time evolution of the ejecta.

\subsection{Progenitor dependence}
\label{sec:prog}

The reverse shock depends on the pressure of the more slowly moving earlier
ejecta, which is different for different progenitor stars, as shown in
Paper~I. For similar explosion energies, more massive progenitors have
slower ejecta and therefore the ejecta shell possesses a higher pressure 
so that the reverse shock stays
at a smaller radius. We have seen in the previous section that an
anisotropic distribution of the pressure in 2D has a big impact on the
reverse shock position. It is thus interesting to study the combined
effect of the two ingredients: progenitor structure and anisotropic
ejecta. 

\begin{figure}[!ht]
  \centering
    \includegraphics[width=0.9\linewidth]{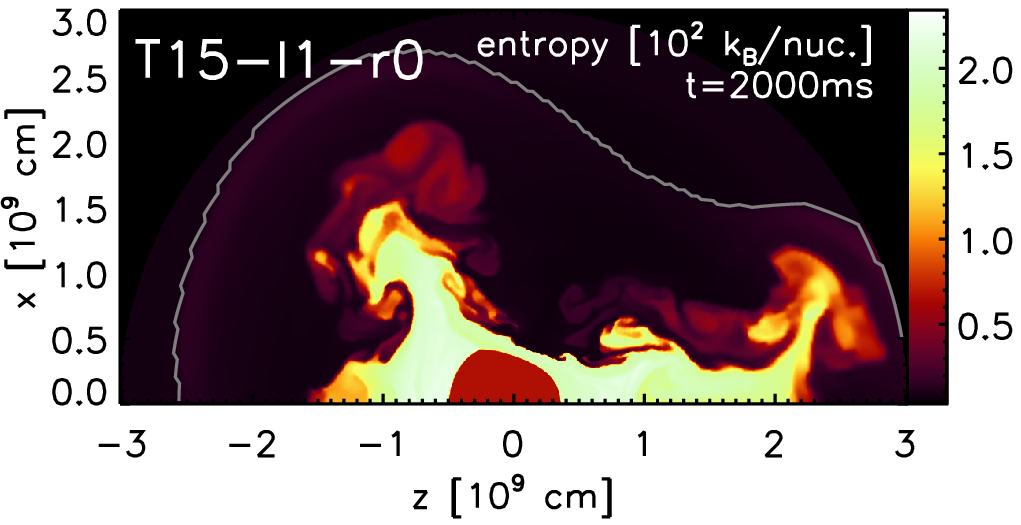} \\
    \includegraphics[width=0.9\linewidth]{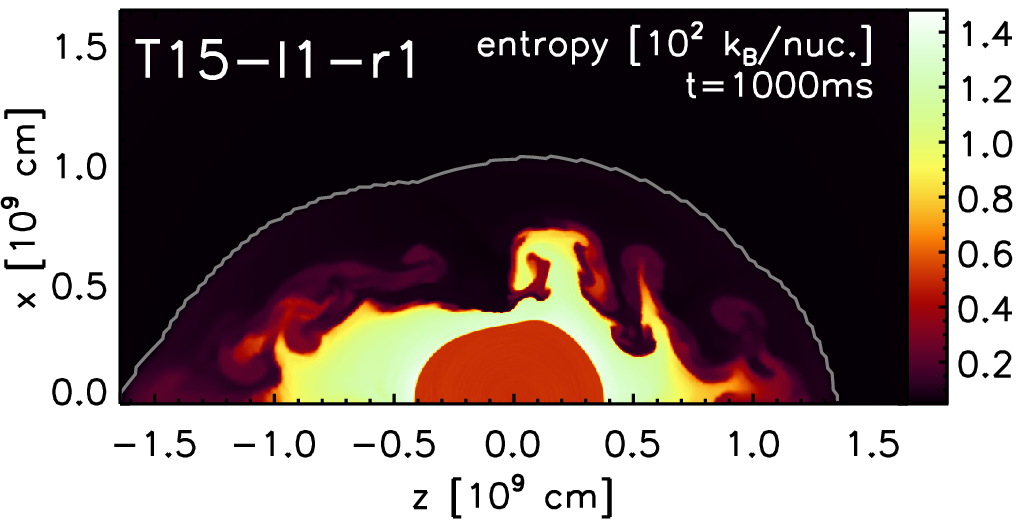} \\
    \includegraphics[width=0.9\linewidth]{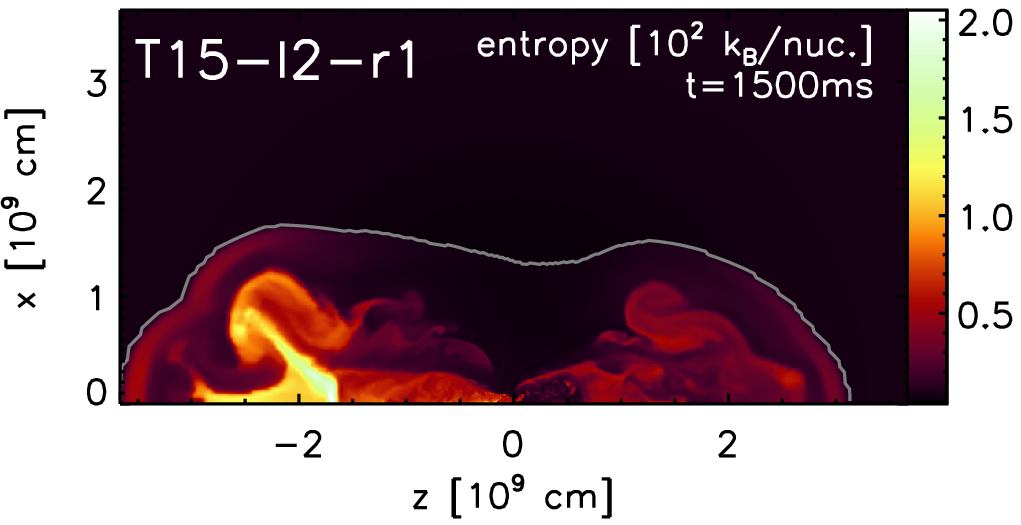} \\
    \includegraphics[width=0.9\linewidth]{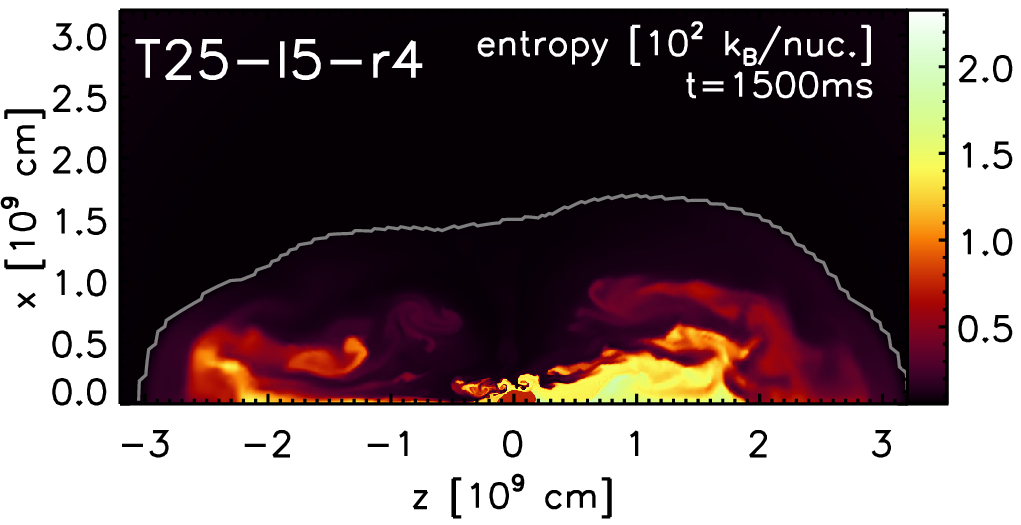} \\
  \caption{Entropy distribution for models T15-l1-r0, T15-l1-r1, and
    T15-l2-r1 of a 15~$M_\odot$ progenitor and model T25-l5-r4 of a 
    25~$M_\odot$ star at the end of the simulations.}
  \label{fig:prog}
\end{figure}

In Fig.~\ref{fig:prog} the entropy distributions are shown for the rest
of the models in Table~\ref{tab:finalresults2d}, representing the
explosions of 15~$M_\odot$ and 25~$M_\odot$ progenitors. Note that the
time is different in every panel because the panels correspond to the 
moments
$t_{\mathrm{end}}$ when our simulations were stopped.  We can compare
models with the same parameters for the proto-neutron star evolution,
but different
progenitors, e.g., T10-l1-r1 and T15-l1-r1.  The evolution of the
ejecta in these two models is considerably different, although the
boundary conditions are the same. This is even more visible in the case
of models T10-l2-r1 and T15-l2-r1. The wind termination shock is almost 
spherically symmetric in the 10~$M_\odot$ star, while it is highly
asymmetric due to the presence of long-lasting downflows in the
15~$M_\odot$ progenitor. The third model for the 15~$M_\odot$ case
(Fig.~\ref{fig:prog}, upper panel) also develops a relatively spherical
reverse shock. Although there is no unambiguous relation between the 
evolution of the ejecta and the shape of the reverse shock on the one
hand and the boundary conditions or the progenitor structure on the
other, more massive progenitor stars like T25-l5-r4 favor long-lasting
and more slowly expanding downdrafts because of their denser
structure and higher mass infall rates. The same trend can be observed
in the cases of lower boundary luminosities (e.g. in T15-l2-r1) and
less energetic explosions. For the explosion models of the 10~$M_\odot$
star shown in Fig.~\ref{fig:T10-stot}, the expansion is relatively fast
because of the more dilute silicon and oxygen shells, and therefore all
downflows are blown away during the first half a second, even when 
lower luminosities are assumed in the simulations.

However, the ejecta distribution does not only depend on the
progenitor and on variations of the boundary conditions that influence
the strength of the explosion. Initial random perturbations must be
imposed on the progenitor (or post-bounce) model to trigger the growth
of instabilities and anisotropies.  These develop in a stochastic and
chaotic way and thus can lead to significant variations of the ejecta
morphology, i.e., of the number and position of downflows and the
direction and size of high-entropy bubbles, all of which have an
influence on the deformation of the forward and reverse shocks and
thus on associated variations of the nucleosynthesis-relevant
conditions in the early supernova ejecta and in the neutrino wind.

In spite of that, for roughly similar explosion energy the progenitor
structure has a systematic influence --at least in a constrained
sense-- on the evolution of the supernova shock and of the
wind-termination shock: As seen from the comparison of model T10-l1-r1
(Fig.~\ref{fig:T10-stot}, left) with models T15-l1-r0, T15-l1-r1, and
T15-l2-r1 (Fig.~\ref{fig:prog}), all of which possess similar
explosion energy, both shocks expand significantly more slowly in the
case of the more massive progenitor. This trend was already obtained
for 1D models in Paper~I and is independent of the explosion
asymmetries that have developed as a consequence of the (stochastic
and chaotic) growth of nonradial instabilities in the shell enclosed
by the forward shock and the reverse shock. Such a systematic
dependence of the ejecta evolution on the progenitor structure is
therefore independent of the dimension and for this reason also of the
particular initial seed perturbation by which the ejecta asymmetry was
initiated to grow in our 2D models. The reason for this
progenitor-dependent variation is the denser structure of the shells
around the iron core in more massive progenitor stars as discussed in
the second paragraph of this section and in more detail in Paper~I.

The multi-dimensional situation, however, allows for an even more
extreme effect that is not possible in the 1D case: In progenitors
significantly more massive than our 10~$M_\odot$ model (or,
alternatively, in the case of very weak explosions), long-lasting
accretion by the neutron star can take place and the explosion can
become extremely asymmetric such that the neutrino-driven outflow can
be confined to narrow angular wedges (see models T15-l2-r1 and
T25-l5-r4 in Fig.~\ref{fig:prog}).  While neutron-star accretion is
still going on, the outflow is then prevented to become supersonic in
some directions so that a termination shock can be absent at these
angles.  It is therefore clear that basic trends defined by progenitor
properties (like those mentioned above) are superimposed in the
multi-dimensional situation by potentially very large nonradial
explosion asymmetries, which cannot be predicted merely on grounds of
the progenitor structure and the value of the explosion energy, but
can have a high relevance for the nucleosynthetic output of the
supernova.

%

\subsection{Discussion of assumptions, approximations, and future 
improvements}
\label{sec:discussion}

In the present work our goal was to identify differences between the
nucleosynthesis-relevant conditions of the neutrino-driven wind in two
dimensions compared to spherical symmetry. We, however, did not intend
to cover the topic of multi-dimensional effects in the winds in its
full breadth, but concentrated in particular on the interaction of the
wind with the more slowly moving, dense ejecta shell behind the
supernova shock. This interaction has obviously stronger and more
important consequences than a variety of other nonradial perturbations
that can make the wind outflow anisotropic at typically much smaller
level.

The wind termination (or reverse) shock, which separates the
supersonic wind region from the subsonic ejecta shell,
exhibits strong non-spherical deformations because of the
large-scale inhomogeneities imprinted on the structure of the ejecta
by the nonradial hydrodynamic instabilities at the beginning of the
explosion. When the reverse shock is located at a sufficiently small
radius, it alters the wind conditions at such a high temperature 
that the nucleosynthesis in the outflow can be strongly affected.
Different from the one-dimensional case, where the reverse-shock
radius is simply a function of the wind properties (mass-loss 
rate and expansion velocity) and explosion properties (energy and
shock radius) as discussed in Paper~I, the location of the reverse
shock in the multi-dimensional case also depends on the inhomogeneities
that are present in the ejecta shell. Dense relics of long-lasting 
accretion funnels, for example, propagate more slowly and prevent
the reverse shock from rapid expansion.

In order to gain insight into such effects, we followed 
two-dimensional models for sufficiently long evolution periods
with the emphasis on maintaining high spatial resolution from 
the steeping density gradient at the proto-neutron star surface 
through the neutrino-wind region, reverse shock, and inhomogeneous
ejecta shell, out to the expanding supernova shock. We employed a
number of assumptions and approximations which we will critically
discuss in the following.

We emphasize that in the present work we want to highlight
basic and general aspects of anisotropic wind-ejecta interaction.
However, we neither intend to make predictions for particular 
progenitors nor do we want to give a statistical overview of 
possible outcomes. Both are not possible in view of our incomplete 
knowledge of the exact explosion mechanism and the use of 
two-dimensional models.

\subsubsection{Random initial seed perturbations}

In order to initiate the growth of nonradial flows in hydrodynamically
unstable regions, we perturbed our spherical initial models with
random seeds (Sect.~\ref{sec:method}).  Random perturbations are a
simple, flexible, and partly tested (see below) procedure to introduce
small-amplitude deviations from spherical symmetry in any available
progenitor or post-bounce model. Of course, it would be interesting to
start with a more realistic assumption based on predictions from
self-consistent stellar evolution modeling. However, multi-dimensional
simulations were performed only for special evolution stages of
selected progenitor stars \cite[e.g.,][ and references
therein]{Young.etal:2005}. Therefore, there are no multi-dimensional
progenitor data at the onset of the gravitational instability for a
range of stellar masses and the fluctuations of quantities in the core
at this stage are not well constrained.

Moreover, tests suggest that the exact properties of the initial
perturbation pattern are not overly important except in the improbable
and special case that a low-mode, global asymmetry is present in the
progenitor core. The influence of different random seeds ---different
amplitudes, different random realizations, and also initial
perturbations of different quantities--- has been explored (at least
to some extent) by
\cite{Buras.etal:2006,Scheck.Kifonidis.Janka.Mueller:2006}. One might
suspect that a high-mode, low-amplitude pattern of the initial
perturbations may have an influence on the fastest growing mode(s) in
the hydrodynamically unstable regions after core bounce as well as on
the linear growth timescales of these most unstable mode(s). However,
it seems unlikely that any memory of such a high-mode seed pattern is
retained once the hydrodynamical instabilities have reached the fully
nonlinear regime because the involved growth is characterized by
highly chaotic and stochastic interactions of structures on different
scales.

\cite{Buras.etal:2006} tested the amplification of random
perturbations on the level of 1--2\% imposed on the density structure
of the progenitor core before collapse. They found that these initial
seeds increase by factors of a few in the supersonically infalling
layers during the collapse (and are damped in the subsonically falling
regions) as predicted by \cite{Lai.Goldreich:2000}. However, the
changes of the model evolution after core bounce turned out to be
insignificant compared to the standard case in which they imposed
perturbations on the level of 0.1\% just after core bounce and thus
only shortly before the hydrodynamically unstable conditions develop
in the supernova core (see Appendix~E in \cite{Buras.etal:2006}).

\cite{Scheck.Kifonidis.Janka.Mueller:2006} investigated the effects of
different random seeds imposed on models after core bounce, perturbing
different quantities ---radial velocity or density--- with an
amplitude of 0.1 percent.  They found that different random
realizations (keeping the progenitor star and boundary conditions the
same) lead to differences of the convective patterns and thus of the
detailed structures of the anisotropies in the dense shell behind the
supernova shock. But while the number of long-lasting downflows, their
locations, the shock deformation and the angular distribution and
volume of buoyant high-entropy bubbles varied, little sensitivity to
the initial seeds was seen for the global properties of the supernova
explosion, i.e., the explosion timescale and the average shock radius
and explosion energy as functions of time. Of course, the transition
from a long-lasting, asymmetric accretion phase to the spherical
neutrino wind will depend on the size and persistance of the
asymmetric accretion funnels that are formed when vigorous, non-radial
instabilities stir the ejecta during the early explosion stage.

For the present paper we picked a set of models that provides a kind
of representative overview of different possible morphologies (see
Figs.~\ref{fig:T10-stot} and \ref{fig:prog}) that were also seen in
the larger set of calculations by
\cite{Scheck.Kifonidis.Janka.Mueller:2006}.

\subsubsection{Anisotropic neutrino emission and flavor conversions}

We describe the neutrino emission from the dense core of the
proto-neutron star by choosing time-dependent (spherically
symmetric) luminosities for
neutrinos and antineutrinos of all flavors at the inner grid
boundary, which is placed significantly below the neutrinosphere.

Naturally, the neutrino flux evolution imposed there has a strong
influence on all stages of the evolution of neutrino-driven
supernovae: on the shock revival and explosion, on the potentially
long-lasting post-explosion accretion, and in particular also on the
power and duration of the neutrino-driven wind.  The dependence of the
dynamical and thermodynamical properties of the neutrino wind
(mass-loss rate, entropy, and expansion timescale) on the luminosities
and mean energies of the radiated neutrinos was in detail discussed in
previous works, in particular by \cite{Qian.Woosley:1996} and by
\cite{Thompson.Burrows.Meyer:2001}, while the dependence of the wind
termination shock on the neutrino-wind properties was the subject of
Paper~I.  Making assumptions about the neutrino fluxes at the inner
grid boundary implies that we can only explore possible dynamical
scenarios in a parameterized way, but we are unable to determine what
exactly happens in a particular progenitor star.

Similarly, the neutrino-flavor evolution necessarily plays a 
crucial role for setting the neutron-to-proton ratio in the wind. 
Thus it decides about the nucleosynthesis processes that can take 
place in this environment. 
While we think that it is not possible to make reliable
predictions of the electron fraction $Y_e$ in the wind on the 
basis of our transport approximations, we are confident that the
set of models presented here can still shed light on fundamental
aspects of anisotropies that can occur during the dynamical 
evolution driven by the neutrino heating between the neutron-star
surface and the supernova shock.

Asymmetric neutrino emission, e.g.\ associated with non-spherical
accretion or convective activity inside the nascent neutron star,
might also play a role for creating anisotropies in the wind ejecta.
The most extreme case of asymmetric accretion are long-lasting
downflows on one side of the proto-neutron star, while mass loss in
the wind is going on in other directions. Such a situation can ---at
least temporarily--- occur in our simulations.  Because of the violent
impact of the accretion flows on the neutron star surface, associated
with rapid motions of the tips of the accretion funnels, strong sonic
waves are launched and create non-radial perturbations in the
neutrino-heated outflows \cite[]{Scheck.Kifonidis.Janka.Mueller:2006}.

These hydrodynamical perturbations of the winds are usually much
bigger than any asymmetries caused by anisotropic neutrino emission
from the accretion downflows. This is so because the supersonically
infalling matter in the downflows has only little time to experience
neutrino losses: It reaches sufficiently high temperatures and
densities only shortly before it impacts on the neutron-star
surface. After deceleration, it quickly spreads sideways around the
surface of the neutron star and settles into the hot accretion layer
surrounding the accretor. While the
anisotropic emission from the accretion flows contributes to the
neutrino luminosity only on the level of some percent, most of this
anisotropic emission occurs in regions where the optical depth for
neutrinos is much lower than unity. The reabsorption of these
neutrinos is therefore very improbable and is unable to cause any
significant asymmetries in the baryonic wind outflow (typically such
asymmetries remain well below one percent). Instead, the wind is
driven by neutrino interactions in regions much closer to the
neutrinosphere, composed of accreted matter that has experienced
efficient mixing upon being dispersed in a turbulent vortex layer. In
addition, the strong surface gravity of the neutron star prevents
asymmetries in the accretion layer so that this layer acts as the base
of a nearly isotropic wind.

A convective region below the neutrinosphere, which is only partly
included in our computational volume (its inner part is omitted with
the excised neutron-star core), also has little impact on wind
anisotropies. Hydrodynamic simulations of the proto-neutron star
formation and of the accretion during the first hundreds of
milliseconds after bounce \cite[]{Buras.Rampp.ea:2006,Dessart06a} show
that the convective shell is placed at a density well above that of
the neutrinosphere. The convective layer is found to be separated from
the neutrinospheric region by a stable shell. The proto-neutron star
convection in the post-explosion phase (later than one second after
bounce) has not been explored by hydrodynamic simulations yet. We
suspect ---and conclude from the structure of our present models near
the inner grid boundary--- that this separation between convective
layer and neutrinosphere is maintained also during the later
stages. Our suspicion is justified by the decreasing optical depth in
the vicinity of the neutrinosphere, which allows neutrino transport to
become more efficient than energy transport by convection. The
convectively stable shell around the convective core of the
proto-neutron star precludes convective asymmetries from having an
immediate and undamped influence on the driving layer of the wind
outflow.  Perturbations of the wind induced by convection thus remain
small.

\cite{Buras.Rampp.ea:2006,Dessart06a} showed that proto-neutron star
convection speeds up the lepton number loss (by temporarily decreasing
the electron-antineutrino luminosity) and increases the energy
transport by higher muon and tau neutrino luminosities.  At the same
time it leads to reduced mean energies of the radiated
neutrinos. While such effects are relevant for detailed modeling of
the Kelvin-Helmholtz cooling of the nascent neutron star, they are
subsumed by our treatment of the inner grid boundary. We therefore do
not expect them to introduce any qualitatively new features into our
models beyond what we can and have explored by variations of the
boundary condition.

Neutrino-flavor conversions near and above the proto-neutron star
surface, in particular due to neutrino-neutrino forward scattering
\cite[see][for a recent review]{Duan.Fuller.Qian:2010} are disregarded
in the present work, but may change the luminosity and spectral
evolution of the different kinds of neutrinos radiated by the forming
neutron star.  During the post-bounce accretion phase and the onset of
the explosion, when the luminosities and spectra of different neutrino
flavors show pronounced differences, such collective neutrino
oscillations might have some relevance for the dynamics and for
setting the neutron-to-proton ratio and thus the nucleosynthesis
conditions in the early supernova ejecta.  However, it is more
uncertain whether any important effects remain during the
Kelvin-Helmholtz cooling of the proto-neutron star after the explosion
has been launched. Hydrodynamical calculations of this phase in
spherical symmetry, using sophisticated neutrino transport methods
(but also ignoring the effects of neutrino flavor oscillations;
\cite{Huedepohl.etal:2010,Fischer.etal:2010}) predict that after the
end of the post-bounce accretion the luminosities and spectra for all
kinds of neutrinos become very similar. Given that, flavor conversions
will certainly have little influence on the neutrino-energy deposition
and therefore on the dynamics of the neutrino-driven wind. Moreover,
also their effect on the electron fraction in the neutrino-heated
outflow is probably relatively minor. Nevertheless, subtle changes of
the neutrino flux spectra due to energy- and direction-dependent
oscillation effects cannot be excluded. Since the neutron-to-proton
ratio and the element formation in the wind outflow are highly
sensitive to details of the electron neutrino and antineutrino
luminosities and spectra, a closer investigation of this aspect may be
desirable, but is very ambitious in the multi-dimensional context.

\subsubsection{General relativity}

The approximation of general relativistic gravity used in our
simulations has been tested in great detail for its accuracy in
different phases of stellar core collapse.  Corresponding results were
published in several preceding papers.
\cite{Marek.Dimmelmeier.ea:2006} compared the neutron star structure,
stability properties, and oscillation behavior obtained with the
effective relativistic potential and fully relativistic calculations
of nonrotating and rotating collapsing stellar iron cores.
\cite{Liebendoerfer.Rampp.ea:2005} compared the dynamics and neutrino
emission during the collapse phase, core bounce, and early post-bounce
evolution in 1D \cite[see
also][]{Marek.Dimmelmeier.ea:2006}. \cite{arcones.janka.scheck:2007}
(Paper I) investigated the structure of the neutrino-driven wind in
comparison to steady-state relativistic wind
solutions. \cite{Huedepohl.etal:2010} and in particular
\cite{Mueller.Janka.Dimmelmeier:2010} explored the Kelvin-Helmholtz
cooling evolution of proto-neutron stars and their neutrino
emission. In these studies fully relativistic models were contrasted
with simulations in which approximated relativistic gravity (plus
gravitational redshift in the neutrino transport where applied) were
combined with otherwise Newtonian hydrodynamics.

The core-collapse phase, core bounce, shock break-out, and early
accretion were found to be in excellent agreement, the neutron-star
structure, its binding energy, neutrino emission properties and total
energy loss were found to be reproduced in all quantities to better
than 10\% and in most cases and phases even to percent accuracy, and
the neutrino-driven wind profiles turned out to be nicely compatible
with steady-state relativistic solutions.  Although there can be
differences in smaller details, the main effects of general relativity
seem to be captured well by our approximative treatment. The reasons
for this result are the facts that (1) the effective relativistic
potential was constructed to explicitly include the relativistic
modifications of the Tolman-Oppenheimer-Volkoff equations compared to
Newtonian hydrostatic equilibrium
\cite[see][]{Marek.Dimmelmeier.ea:2006}, and (2) the fluid velocities
in the supernova core hardly ever exceed 10--20\% of the speed of
light, in which case Newtonian fluid dynamics is still a good
approximation.

\subsubsection{Neutron star motion}

A kicked neutron star will move away from the grid center but will
remain the center of the neutrino-driven wind. The neutron star motion
may thus lead to a distortion of the wind-ejecta interaction and may
change the location of the reverse shock in the wind and thus its
influence on the wind properties. Such influence on the detailed
evolution of a particular model was indeed seen by
\cite{Scheck.Kifonidis.Janka.Mueller:2006}, who performed runs (for
fixed progenitor, boundary conditions, and perturbation seeds) without
and with applying a Galilei transformation on the flow around a
neutron star at the grid center by giving the surrounding fluid a
coherent motion with the negative sign of the neutron star kick
velocity. This procedure was assumed to capture the main effects of
the neutron star movement.

Although such effects can be relevant for determining the detailed
evolution and ejecta properties of a particular model star 
(which was not the goal of this work)
they are unlikely to change the basic and general conclusions that
we have drawn from our simulations. Since the neutron-star velocity
$v_\mathrm{ns}$ is typically small compared to the sound speed and
the terminal wind velocity, $v_\mathrm{ns} \ll c_\mathrm{s}$ and
$v_\mathrm{ns} \ll u_\mathrm{w}(r \gg R_\mathrm{ns})$, it is
hard to imagine that allowing the neutron star to move (instead
of anchoring it at the grid center as in our models) will affect
the overall picture that we have obtained for the anisotropic
wind-ejecta interaction.

\subsubsection{Three-dimensional modeling}

Performing simulations in two dimensions imposes the artifical
constraint that all structures are axially symmetric. Of course,
this naturally raises the question what differences one might
expect in three dimensions.

Ignoring the important question whether 2D/3D differences have any
influence on the success of the neutrino-heating mechanism for
supernova explosions (a very first statement in this context was
recently published by \cite{Nordhaus.Burrows.etal:2010}), and relying
on the viability of this mechanism (which is a fundamental assumption
in our study), basic features of the 2D asymmetries observed in our
calculations seem to be confirmed by recent 3D simulations:
\cite{Wongwathanarat.Janka.Mueller:2010} followed the evolution of a
small set of 3D models over a similarly long post-bounce period (but
with considerably less spatial resolution than used in our 2D
calculations).  The 3D slice of figure~3, right panel, in the latter
publication exhibits the same overall behavior and structural features
for the neutrino-driven wind phase in 2D and 3D models, namely an
essentially spherical wind, an asymmetric wind-termination shock, and
an inhomogeneous and strongly anisotropic ejecta shell behind the
forward shock. Also in 3D the reverse shock is highly deformed and its
effect on the wind properties and nucleosynthesis conditions depends
on the outflow direction.  We are therefore confident that our main
findings for the wind-ejecta interaction are valid not only in the
considered 2D situation.

Of course, the present parameterized explosion models, which
do not yield any information about the explosion properties
(energy, timescale) of a progenitor star, cannot be conclusive
on the consequences of these asymmetries for the explosive
nucleosynthesis of individual stars, nor does our present 
knowledge of the explosion mechanism allow for any statements
in a statistical sense.

Rotation of the nascent neutron star is an additional degree
of freedom, which we ignored in the models
of this paper. It will cause a global pole-equator asymmetry of
the wind and of the wind-ejecta interaction with potentially 
interesting implications. This should be studied in future work
by systematic variations of the proto-neutron star spin.

\section{Nucleosynthesis implications}
\label{sec:nuc}

We have shown the impact of multidimensional effects on the dynamical
evolution of the neutrino-driven wind, reverse shock and supernova
ejecta. In this section, we want to briefly address the possible
implications of our results for the nucleosynthesis processes
occurring in supernova outflows: charged-particle reactions, alpha
process \cite[]{Woosley.Hoffman:1992, Witti.Janka.Takahashi:1994},
$\nu$p-process \cite[]{Froehlich.Martinez-Pinedo.ea:2006,
  Pruet.Hoffman.ea:2006, Wanajo:2006}, and occasionally r-process
\cite[][for a review]{arnould.goriely.takahashi:2007}.

Since the works of \cite{Cameron:1957} and
\cite{Burbidge.Burbidge.ea:1957}, core-collapse supernova outflows
have been the best studied candidate for the production of heavy
elements. However, this environment is facing more and more problems
to fulfill the requirements (high entropy, low electron fraction and
fast expansion) for the production of heavy r-process elements
(A$>90$). The conditions found to be necessary 
for a robust and strong r-process \cite[e.g.,][]{hoffman.woosley.qian:1997}
are not achieved by recent long-time supernova simulations
\cite[Paper~I,][]{Huedepohl.etal:2010, Fischer.etal:2010}. This is also
the case for our 2D simulations, where the wind entropies are too low
to get the high neutron-to-seed ratios necessary for the r-process.

Yet galactic chemical evolution models \cite[see
e.g.,][]{Ishimaru.etal:2004, Qian.Wasserburg:2008} suggest that at
least a subset of core-collapse supernovae could be responsible of the
origin of half of the heavy r-process elements. Therefore, one may
speculate that the r-process could take place in neutrino-driven winds
because of some still unknown aspect of physics that might cure the
problems revealed by the present hydrodynamical models.  In this case
the reverse shock could have important consequences
\cite[]{Wanajo.Itoh.ea:2002}. Depending on the temperature at the
reverse shock the r-process path is different.  When the reverse-shock
temperature is low ($T_{\mathrm{rs}} \lesssim 0.5$~GK),
neutron-capture and beta-decay timescales are similar
\cite[]{Blake.Schramm:1976} and shorter than $(\gamma,n)$
timescales. This is also known as ``cold r-process''
\cite[]{Wanajo:2007, Panov.Janka:2009}. In contrast, when the reverse
shock is at high temperatures, there is an $(n,\gamma)-(\gamma,n)$
equilibrium as in the classical r-process
\cite[]{Kratz.Bitouzet.ea:1993, Freiburghaus.Rembges.ea:1999,
  Farouqi.etal:2010}.  The final abundances for these two types of
evolution are very different, see e.g., \cite{Wanajo.Itoh.ea:2002,
  Wanajo:2007, Kuroda.Wanajo.Nomoto:2008, Panov.Janka:2009,
  Arcones.Martinez-Pinedo:2010}.  Also the impact of the nuclear
physics input varies depending on how matter expands
\cite[]{Arcones.Martinez-Pinedo:2010}.

In addition to the r-process, whose astrophysical site 
is still uncertain, there are
other nucleosynthesis processes occurring in supernova outflows. The
reverse shock can affect the production of p-nuclei that happens in
neutrino-driven winds via charged-particle reactions
\cite[]{Woosley.Hoffman:1992} and the $\nu$p-process
\cite[]{Froehlich.Martinez-Pinedo.ea:2006, Pruet.Hoffman.ea:2006,
  Wanajo:2006}. This is becoming more important, because the most
recent and most sophisticated supernova simulations show that the
ejecta are proton-rich for several seconds
\cite[]{Fischer.etal:2010} and even until completion of the
proto-neutron star cooling and deleptonization
\cite[]{Huedepohl.etal:2010}. The impact of the reverse shock on these
nucleosynthesis processes can show up in two ways.  First, the
temperature jump leads to an increase of the photo-dissociation
rate. When the latter is too high, newly formed nuclei are
destroyed. However, when the photo-dissociation rate is moderate, the
temperature increase favors the captures of charged particles.  The
other important effect of the reverse shock is the strong reduction of
the expansion velocity with the consequence that the temperature stays
constant or decreases only slowly after a mass element has crossed the
reverse shock. Depending on the exact value of the temperature,
photo-dissociations or charged-particle reactions continue to take
place. Moreover, when the expansion becomes slower, the matter stays
exposed to high neutrino fluxes for a longer time. This increases the
efficiency of the $\nu$p-process. However, one should notice that the
processes described here are possible only when the reverse-shock
radius is sufficiently small, e.g., during the first few seconds after
the onset of the explosion, because otherwise the temperature at the
reverse shock is already too low to play any role. Recently,
\cite{Wanajo.etal:2010} suggested a possible significant impact of the
reverse shock on the nucleosynthesis under these conditions, in
contrast to the small effects reported by
\cite{Roberts.etal:2010}. Therefore, further nucleosynthesis studies
should be done, taking into account the reverse-shock behaviour found
in Paper~I and in particular in the present work, where a huge
variability due to multi-dimensional effects was obtained.

\section{Conclusions}
\label{sec:conclusions}

With a small set of two-dimensional simulations for three progenitor
stars of different masses (Table~\ref{tab:finalresults2d})
we have demonstrated that the reverse-shock
radius and the conditions of the shocked neutrino-driven wind
matter in supernova explosions become strongly angle dependent. As we
found in Paper~I by analytic means, the position of the reverse shock
depends on the wind properties (mass outflow rate and velocity) as
well as on the pressure of the more slowly moving supernova ejecta.
Comparison of the radial profiles of 1D and 2D simulations shows that the
neutrino-driven wind is spherically symmetric. This is caused by the
isotropic neutrino emission from a neutron star that stays spherical
in the absence of rotation. Multi-dimensional simulations including
rotation could lead to differences in the wind
\cite[]{Metzger.etal:2006, Wanajo06} and thus significant changes in
its interaction with the supernova ejecta. Although without rotation
the wind develops identically in all directions, a strong angular
variation of the reverse shock position can appear because of the
anisotropic matter (and thus pressure) distribution in the more
slowly moving, early supernova ejecta.

When the radial location of the reverse shock is constrained by the 
existence of dense downdrafts in the ejecta shell that follows
the supernova shock, the angle between the reverse shock and the
wind velocity, which goes in the radial direction, can be estimated.
We have found an analytic expression that relates this angle to the
jump of the pressure at the reverse shock. The presence of the
downflow features in the supernova ejecta shell with local density
and pressure maxima leads to angular variations of the radius of
the reverse shock and thus a deformation of the wind-shell boundary
with kinks appearing in the regions where the
obliqueness of the shock abruptly changes.  The Rankine-Hugoniot
conditions imply that oblique shocks are less efficient in
decelerating matter. This produces a collimation of the shocked flow
in the vicinity of the kinks.  Finally, we have proven that basic
features of the progenitor dependence seen in Paper~I are also present
in our two-dimensional simulations, e.g., a slower expansion of the
forward and reverse shocks in more massive stars. However, such
general aspects are superimposed by an enormous amount of variability
of explosions even of the same progenitor and similar explosion energy
due to the chaotic growth of nonradial hydrodynamic instabilities from
small initial seed perturbations.

In summary, we have found that in the multi-dimensional case the
expansion of the wind matter varies with the angular direction because
it is influenced by the anisotropic distribution of the earlier
ejecta, which evolves chaotically from initial random
perturbations. Therefore, we strongly recommend that future
nucleosynthesis studies should test the effect of different
extrapolations of the evolution of the shock-decelerated ejecta. Our
results can be used as guidance for the overall variability that is
possible and affects the nucleosynthesis-relevant conditions in
multi-dimensional supernova environments.


\begin{acknowledgements}
  We thank G.~Mart\'inez-Pinedo, F.~Montes, I.~Panov,
  F.-K.~Thielemann, and S.~Wanajo for stimulating discussions.  We are
  grateful to S.~Woosley and A.~Heger for providing us with the
  progenitor models, to A.~Marek for computing the collapse and prompt
  shock propagation phases with the \textsc{Vertex-Prometheus}
  neutrino-hydrodynamics code.  A.~Arcones is supported by the Swiss
  National Science Foundation.  The work of H.-T.~Janka and
  collaboration visits were supported by the Deutsche
  Forschungsgemeinschaft through the Transregional Collaborative
  Research Centers SFB/TR~27 ``Neutrinos and Beyond'' and SFB/TR~7
  ``Gravitational Wave Astronomy'', and the Cluster of Excellence
  EXC~153 ``Origin and Structure of the Universe''
  (http://www.universe-cluster.de).  The computations were performed
  on the NEC SX-5/3C and the IBM p690 ``Regatta'' system of the
  Rechenzentrum Garching, and on the IBM p690 cluster ``Jump'' of the
  John von Neumann Institute for Computing (NIC) in J\"ulich.
\end{acknowledgements}


\begin{thebibliography}{58}
\expandafter\ifx\csname natexlab\endcsname\relax\def\natexlab#1{#1}\fi

\bibitem[{{Arcones} {et~al.}(2007){Arcones}, {Janka}, \&
  {Scheck}}]{arcones.janka.scheck:2007}
{Arcones}, A., {Janka}, H.-T., \& {Scheck}, L. 2007, \aap, 467, 1227

\bibitem[{{Arcones} \& {Martinez-Pinedo}(2010)}]{Arcones.Martinez-Pinedo:2010}
{Arcones}, A. \& {Martinez-Pinedo}, G. 2010, submitted to Phys.~Rev.~C, arXiv:1008.3890

\bibitem[{{Arcones} \& {Montes}(2010)}]{Arcones.Montes:2010}
{Arcones}, A. \& {Montes}, F. 2010, submitted to ApJ, arXiv:1007.1275

\bibitem[{Arnould {et~al.}(2007)Arnould, Goriely, \&
  Takahashi}]{arnould.goriely.takahashi:2007}
Arnould, M., Goriely, S., \& Takahashi, K. 2007, Phys. Repts., 450, 97

\bibitem[{{Blake} \& {Schramm}(1976)}]{Blake.Schramm:1976}
{Blake}, J.~B. \& {Schramm}, D.~N. 1976, \apj, 209, 846

\bibitem[{{Buras} {et~al.}(2006){Buras}, {Janka}, {Rampp}, \&
  {Kifonidis}}]{Buras.etal:2006}
{Buras}, R., {Janka}, H.-T., {Rampp}, M., \& {Kifonidis}, K. 2006, \aap, 457,
  281

\bibitem[{Buras {et~al.}(2006)Buras, Rampp, Janka, \&
  Kifonidis}]{Buras.Rampp.ea:2006}
Buras, R., Rampp, M., Janka, H.-T., \& Kifonidis, K. 2006, \aap, 447, 1049

\bibitem[{{Burbidge} {et~al.}(1957){Burbidge}, {Burbidge}, {Fowler}, \&
  {Hoyle}}]{Burbidge.Burbidge.ea:1957}
{Burbidge}, E.~M., {Burbidge}, G.~R., {Fowler}, W.~A., \& {Hoyle}, F. 1957,
  Rev. Mod. Phys., 29, 547

\bibitem[{Burrows {et~al.}(1995)Burrows, Hayes, \&
  Fryxell}]{Burrows.Hayes.Fryxell:1995}
Burrows, A., Hayes, J., \& Fryxell, B.~A. 1995, \apj, 450, 830

\bibitem[{Cameron(1957)}]{Cameron:1957}
Cameron, A. G.~W. 1957, Stellar Evolution, Nuclear Astrophysics, and
  Nucleogenesis, Report CRL-41, Chalk River

\bibitem[{{Dessart} {et~al.}(2006){Dessart}, {Burrows}, {Livne}, \&
  {Ott}}]{Dessart06a}
{Dessart}, L., {Burrows}, A., {Livne}, E., \& {Ott}, C.~D. 2006, \apj, 645, 534

\bibitem[{{Duan} {et~al.}(2010){Duan}, {Fuller}, \&
  {Qian}}]{Duan.Fuller.Qian:2010}
{Duan}, H., {Fuller}, G.~M., \& {Qian}, Y. 2010, Annual Review of Nuclear and
  Particle Science, 60, 569

\bibitem[{{Duncan} {et~al.}(1986){Duncan}, {Shapiro}, \&
  {Wasserman}}]{duncan.shapiro.wasserman:1986}
{Duncan}, R.~C., {Shapiro}, S.~L., \& {Wasserman}, I. 1986, \apj, 309, 141

\bibitem[{{Farouqi} {et~al.}(2010){Farouqi}, {Kratz}, {Pfeiffer}, {Rauscher},
  {Thielemann}, \& {Truran}}]{Farouqi.etal:2010}
{Farouqi}, K., {Kratz}, K., {Pfeiffer}, B., {et~al.} 2010, 712, 1359

\bibitem[{{Fischer} {et~al.}(2010){Fischer}, {Whitehouse}, {Mezzacappa},
  {Thielemann}, \& {Liebend{\"o}rfer}}]{Fischer.etal:2010}
{Fischer}, T., {Whitehouse}, S.~C., {Mezzacappa}, A., {Thielemann}, F., \&
  {Liebend{\"o}rfer}, M. 2010, \aap, 517, A80+

\bibitem[{{Freiburghaus} {et~al.}(1999){Freiburghaus}, {Rembges}, {Rauscher},
  {Kolbe}, {Thielemann}, {Kratz}, {Pfeiffer}, \&
  {Cowan}}]{Freiburghaus.Rembges.ea:1999}
{Freiburghaus}, C., {Rembges}, J.-F., {Rauscher}, T., {et~al.} 1999, \apj, 516,
  381

\bibitem[{Fr{\"o}hlich {et~al.}(2006)Fr{\"o}hlich, Mart{\'\i}nez-Pinedo,
  Liebend{\"o}rfer, Thielemann, Bravo, Hix, Langanke, \&
  Zinner}]{Froehlich.Martinez-Pinedo.ea:2006}
Fr{\"o}hlich, C., Mart{\'\i}nez-Pinedo, G., Liebend{\"o}rfer, M., {et~al.}
  2006, \prl, 96, 142502

\bibitem[{{Heger} {et~al.}(2001){Heger}, {Woosley}, {Mart{\'{\i}}nez-Pinedo},
  \& {Langanke}}]{Heger01}
{Heger}, A., {Woosley}, S.~E., {Mart{\'{\i}}nez-Pinedo}, G., \& {Langanke}, K.
  2001, \apj, 560, 307

\bibitem[{{Hoffman} {et~al.}(1997){Hoffman}, {Woosley}, \&
  {Qian}}]{hoffman.woosley.qian:1997}
{Hoffman}, R.~D., {Woosley}, S.~E., \& {Qian}, Y.-Z. 1997, \apj, 482, 951

\bibitem[{{H{\"u}depohl} {et~al.}(2010){H{\"u}depohl}, {M{\"u}ller}, {Janka},
  {Marek}, \& {Raffelt}}]{Huedepohl.etal:2010}
{H{\"u}depohl}, L., {M{\"u}ller}, B., {Janka}, H., {Marek}, A., \& {Raffelt},
  G.~G. 2010, \prl, 104, 251101

\bibitem[{{Ishimaru} {et~al.}(2004){Ishimaru}, {Wanajo}, {Aoki}, \&
  {Ryan}}]{Ishimaru.etal:2004}
{Ishimaru}, Y., {Wanajo}, S., {Aoki}, W., \& {Ryan}, S.~G. 2004, \apjl, 600,
  L47

\bibitem[{Janka {et~al.}(2007)Janka, Langanke, Marek, Mart{\'i}nez-Pinedo, \&
  M{\"u}ller}]{Janka.Langanke.ea:2007}
Janka, H.-T., Langanke, K., Marek, A., Mart{\'i}nez-Pinedo, G., \& M{\"u}ller,
  B. 2007, Phys. Repts., 442, 38

\bibitem[{{Janka} \& {M\"uller}(1995)}]{Janka95}
{Janka}, H.-T. \& {M\"uller}, E. 1995, \apjl, 448, L109

\bibitem[{Janka \& M{\"u}ller(1996)}]{Janka.Mueller:1996}
Janka, H.-T. \& M{\"u}ller, E. 1996, \aap, 306, 167

\bibitem[{{Kifonidis} {et~al.}(2003){Kifonidis}, {Plewa}, {Janka}, \&
  {M{\"u}ller}}]{Kifonidis03}
{Kifonidis}, K., {Plewa}, T., {Janka}, H.-T., \& {M{\"u}ller}, E. 2003, \aap,
  408, 621

\bibitem[{{Kratz} {et~al.}(1993){Kratz}, {Bitouzet}, {Thielemann}, {Moeller},
  \& {Pfeiffer}}]{Kratz.Bitouzet.ea:1993}
{Kratz}, K., {Bitouzet}, J., {Thielemann}, F., {Moeller}, P., \& {Pfeiffer}, B.
  1993, \apj, 403, 216

\bibitem[{{Kuroda} {et~al.}(2008){Kuroda}, {Wanajo}, \&
  {Nomoto}}]{Kuroda.Wanajo.Nomoto:2008}
{Kuroda}, T., {Wanajo}, S., \& {Nomoto}, K. 2008, \apj, 672, 1068

\bibitem[{{Lai} \& {Goldreich}(2000){Kuroda}}]{Lai.Goldreich:2000}
{Lai}, D., \& {Goldreich}, P. 2000, \apj, 535, 402

\bibitem[{{Landau} \& {Lifshitz}(1959)}]{Landau}
{Landau}, L. \& {Lifshitz}, E. 1959, Course of Theoretical Physics. Fluid
  mechanics, Vol.~6 (Pergamon)

\bibitem[{Liebend{\"o}rfer {et~al.}(2005)Liebend{\"o}rfer, Rampp, Janka, \&
  Mezzacappa}]{Liebendoerfer.Rampp.ea:2005}
Liebend{\"o}rfer, M., Rampp, M., Janka, H.-T., \& Mezzacappa, A. 2005, \apj,
  620, 840

\bibitem[{{Marek} {et~al.}(2006){Marek}, {Dimmelmeier}, {Janka}, {M{\"u}ller},
  \& {Buras}}]{Marek.Dimmelmeier.ea:2006}
{Marek}, A., {Dimmelmeier}, H., {Janka}, H.-T., {M{\"u}ller}, E., \& {Buras},
  R. 2006, \aap, 445, 273

\bibitem[{{Metzger} {et~al.}(2007){Metzger}, {Thompson}, \&
  {Quataert}}]{Metzger.etal:2006}
{Metzger}, B.~D., {Thompson}, T.~A., \& {Quataert}, E. 2007, \apj, 659, 561

\bibitem[{{M{\"u}ller} {et~al.}(2010){M{\"u}ller}, {Janka}, \&
  {Dimmelmeier}}]{Mueller.Janka.Dimmelmeier:2010}
{M{\"u}ller}, B., {Janka}, H., \& {Dimmelmeier}, H. 2010, \apjs, 189, 104

\bibitem[{{Nordhaus} {et~al.}(2010){Nordhaus}, {Burrows}, {Almgren}, \&
  {Bell}}]{Nordhaus.Burrows.etal:2010}
{Nordhaus}, J., {Burrows}, A., {Almgren}, A., \& {Bell}, J. 2010,
  arXiv:1006.3792

\bibitem[{{Otsuki} {et~al.}(2000){Otsuki}, {Tagoshi}, {Kajino}, \&
  {Wanajo}}]{Otsuki.Tagoshi.ea:2000}
{Otsuki}, K., {Tagoshi}, H., {Kajino}, T., \& {Wanajo}, S. 2000, \apj, 533, 424

\bibitem[{{Panov} \& {Janka}(2009)}]{Panov.Janka:2009}
{Panov}, I.~V. \& {Janka}, H.-T. 2009, \aap, 494, 829

\bibitem[{{Pons} {et~al.}(1999){Pons}, {Reddy}, {Prakash}, {Lattimer}, \&
  {Miralles}}]{Pons.Reddy.ea:1999}
{Pons}, J.~A., {Reddy}, S., {Prakash}, M., {Lattimer}, J.~M., \& {Miralles},
  J.~A. 1999, \apj, 513, 780

\bibitem[{Pruet {et~al.}(2006)Pruet, Hoffman, Woosley, Janka, \&
  Buras}]{Pruet.Hoffman.ea:2006}
Pruet, J., Hoffman, R.~D., Woosley, S.~E., Janka, H.-T., \& Buras, R. 2006,
  \apj, 644, 1028

\bibitem[{{Qian} \& {Wasserburg}(2008)}]{Qian.Wasserburg:2008}
{Qian}, Y. \& {Wasserburg}, G.~J. 2008, \apj, 687, 272

\bibitem[{{Qian} \& {Woosley}(1996)}]{Qian.Woosley:1996}
{Qian}, Y.-Z. \& {Woosley}, S.~E. 1996, \apj, 471, 331

\bibitem[{{Roberts} {et~al.}(2010){Roberts}, {Woosley}, \&
  {Hoffman}}]{Roberts.etal:2010}
{Roberts}, L.~F., {Woosley}, S.~E., \& {Hoffman}, R.~D. 2010,
  arXiv:1004.4916

\bibitem[{{Scheck} {et~al.}(2006){Scheck}, {Kifonidis}, {Janka}, \&
  {M{\"u}ller}}]{Scheck.Kifonidis.Janka.Mueller:2006}
{Scheck}, L., {Kifonidis}, K., {Janka}, H.-T., \& {M{\"u}ller}, E. 2006, \aap,
  457, 963

\bibitem[{{Sneden} {et~al.}(2008){Sneden}, {Cowan}, \&
  {Gallino}}]{Sneden.etal:2008}
{Sneden}, C., {Cowan}, J.~J., \& {Gallino}, R. 2008, \araa, 46, 241

\bibitem[{{Sumiyoshi} {et~al.}(2000){Sumiyoshi}, {Suzuki}, {Otsuki},
  {Terasawa}, \& {Yamada}}]{Sumiyoshi00}
{Sumiyoshi}, K., {Suzuki}, H., {Otsuki}, K., {Terasawa}, M., \& {Yamada}, S.
  2000, \pasj, 52, 601

\bibitem[{Takahashi {et~al.}(1994)Takahashi, Witti, \&
  Janka}]{Takahashi.Witti.Janka:1994}
Takahashi, K., Witti, J., \& Janka, H.-T. 1994, \aap, 286, 857

\bibitem[{{Terasawa} {et~al.}(2002){Terasawa}, {Sumiyoshi}, {Yamada}, {Suzuki},
  \& {Kajino}}]{Terasawa.Sumiyoshi.ea:2002}
{Terasawa}, M., {Sumiyoshi}, K., {Yamada}, S., {Suzuki}, H., \& {Kajino}, T.
  2002, \apj, 578, L137

\bibitem[{Thompson {et~al.}(2001)Thompson, Burrows, \&
  Meyer}]{Thompson.Burrows.Meyer:2001}
Thompson, T.~A., Burrows, A., \& Meyer, B.~S. 2001, \apj, 562, 887

\bibitem[{Wanajo(2006)}]{Wanajo:2006}
Wanajo, S. 2006, \apj, 647, 1323

\bibitem[{{Wanajo}(2006)}]{Wanajo06}
{Wanajo}, S. 2006, \apjl, 650, L79

\bibitem[{{Wanajo}(2007)}]{Wanajo:2007}
{Wanajo}, S. 2007, \apjl, 666, L77

\bibitem[{{Wanajo} {et~al.}(2002){Wanajo}, {Itoh}, {Ishimaru}, {Nozawa}, \&
  {Beers}}]{Wanajo.Itoh.ea:2002}
{Wanajo}, S., {Itoh}, N., {Ishimaru}, Y., {Nozawa}, S., \& {Beers}, T.~C. 2002,
  \apj, 577, 853

\bibitem[{{Wanajo} {et~al.}(2010){Wanajo}, {Janka}, \&
  {Kubono}}]{Wanajo.etal:2010}
{Wanajo}, S., {Janka}, H., \& {Kubono}, S. 2010, arXiv:1004.4487

\bibitem[{Witti {et~al.}(1994)Witti, Janka, \&
  Takahashi}]{Witti.Janka.Takahashi:1994}
Witti, J., Janka, H.-T., \& Takahashi, K. 1994, \aap, 286, 841

\bibitem[{{Wongwathanarat} {et~al.}(2010){Wongwathanarat}, {Janka}, \&
  {Mueller}}]{Wongwathanarat.Janka.Mueller:2010}
{Wongwathanarat}, A., {Janka}, H., \& {Mueller}, E. 2010, arXiv:1010.0167

\bibitem[{Woosley \& Hoffman(1992)}]{Woosley.Hoffman:1992}
Woosley, S.~E. \& Hoffman, R.~D. 1992, \apj, 395, 202

\bibitem[{Woosley \& Weaver(1995)}]{Woosley.Weaver:1995}
Woosley, S.~E. \& Weaver, T.~A. 1995, \apjs, 101, 181

\bibitem[{Woosley {et~al.}(1994)Woosley, Wilson, Mathews, Hoffman, \&
  Meyer}]{Woosley.Wilson.ea:1994}
Woosley, S.~E., Wilson, J.~R., Mathews, G.~J., Hoffman, R.~D., \& Meyer, B.~S.
  1994, \apj, 433, 229

\bibitem[{{Young} {et~al.}(2005){Young}, {Meakin}, {Arnett}, \&
  {Fryer}}]{Young.etal:2005}
{Young}, P.~A., {Meakin}, C., {Arnett}, D., \& {Fryer}, C.~L. 2005, \apjl, 629,
  L101

\end{thebibliography}

\end{document}